\documentclass[12pt]{iopart}
\usepackage{iopams}
\usepackage{graphicx}
\usepackage{relsize}

\def\dif{\mathrm{d}}
\def\mbr{\mathbf{r}}
\def\mma{\textit{\textbf{a}}}
\def\mA{\mathrm{A}}
\def\mB{\mathrm{B}}
\def\hatd#1{\hat{#1}^\dagger}
\def\bra#1{\left\langle{#1}\right|}
\def\ket#1{\left|{#1}\right\rangle}
\def\braket#1#2{\left\langle{{#1}}\mathrel{\left|{\vphantom{{#1}{#2}}}\right.\kern-\nulldelimiterspace}{{#2}}\right\rangle}
\def\slfrac#1#2{\left.#1\middle/#2\right.}
\def\comma{\mathsmaller{,}}

\begin{document}

\title[]{Topological invariant and cotranslational symmetry in strongly interacting multi-magnon systems}

\author{Xizhou Qin$^{1}$, Feng Mei$^{1}$, Yongguan Ke$^{1,2}$, Li Zhang$^{1,2}$, and Chaohong Lee$^{1,2,*}$}

\address{$^{1}$ TianQin Research Center \& School of Physics and Astronomy, Sun Yat-Sen University (Zhuhai Campus), Zhuhai 519082, China}
\address{$^{2}$ Key Laboratory of Optoelectronic Materials and Technologies, Sun Yat-Sen University (Guangzhou Campus), Guangzhou 510275, China}

\ead{lichaoh2@mail.sysu.edu.cn, chleecn@gmail.com}

\vspace{10pt}
\begin{indented}
\item[]\today
\end{indented}

\begin{abstract}
It is still an outstanding challenge to characterize and understand the topological features of strongly interacting states such as bound-states in interacting quantum systems.
Here, by introducing a cotranslational symmetry in an interacting multi-particle quantum system, we systematically develop a method to define a Chern invariant, which is a generalization of the well-known Thouless-Kohmoto-Nightingale-den Nijs invariant, for identifying strongly interacting topological states.
As an example, we study the topological multi-magnon states in a generalized Heisenberg XXZ model, which can be realized by the currently available experiment techniques of cold atoms [Phys. Rev. Lett. \textbf{111}, 185301 (2013); Phys. Rev. Lett. \textbf{111}, 185302 (2013)].
Through calculating the two-magnon excitation spectrum and the defined Chern number, we explore the emergence of topological edge bound-states and give their topological phase diagram.
We also analytically derive an effective single-particle Hofstadter superlattice model for a better understanding of the topological bound-states.
Our results not only provide a new approach to defining a topological invariant for interacting multi-particle systems, but also give insights into the characterization and understanding of strongly interacting topological states.
\end{abstract}

\vspace{2pc}

\section{Introduction}

Topological invariants, which describe the invariant property of a topological space under homeomorphisms, are of great importance in characterizing topological matters and topological phase transitions.
Weakly interacting topological states, whose universal properties do not depend on inter-particle interactions, are well-understood due to the well-developed tools for treating weakly interacting systems~\cite{KaneRev2010,QiRev2011,Schnyder2008,Kitaev2009,MooreNature2010}.
However, strongly interacting topological states, whose universal properties are determined by inter-particle interactions, pose much greater challenges to both theory~\cite{FTI2015} and experiment~\cite{Grusdt2015}.
The characterization of strongly interacting topological states is quite different from that of the weakly interacting counterparts~\cite{Chen2012,Wang2014}.
Due to the existence of strong correlations among particles, it is hard to define a topological invariant and to clarify the interplay between topological features and inter-particle interactions.

Ultracold atoms in optical lattices offer a well-controlled experimental platform to explore topological matters in a clean environment~\cite{NatPhys-Review}.
Recently, the Hofstadter-Harper model has been experimentally realized by using laser-assisted tunneling of ultracold atoms in a tilted optical potential~\cite{Bloch2013,Ketterle2013}.
As the atom-atom interaction can be tuned by Feshbach resonances, such an atomic Hofstadter-Harper system not only opens a way to explore topological states of noninteracting atoms, but also provides new opportunity to study strongly interacting topological states.

Beyond single-particle topological states~\cite{MH2010,ExpMH2010,Li2013,Pereiro2014,Sachdev2014,Ong2015,Lee2015}, it is of great challenge to clarify whether interacting topological states may emerge.
One outstanding challenge is the absence of a well-defined topological invariant for an interacting quantum system (IQS).
In this paper, we find that this problem can be solved when the system has \emph{cotranslational symmetry}: the invariance under collective translation.
We demonstrate that the cotranslational symmetry naturally allows us to formulate a topological invariant, which can be used to characterize the topological features of interacting multi-particle states such as bound-states (BS's).
In comparison with other topological invariants, our topological invariant is intrinsic and straightforward.
A well-known generalization of the Thouless-Kohmoto-Nightingale-den Nijs (TKNN) invariant~\cite{TKNN1982} from noninteracting to interacting systems is by introducing the twisted boundary condition (BC)~\cite{Niu1985}, which requires to calculate all many-body ground-states for a continuous $2\pi$-period of the twist angle.
Another topological invariant for IQS's is given in terms of the Green's function, which requires to calculate the Green's function at all frequencies~\cite{WangPRL2010} or zero frequency~\cite{WangPRX2012}.
Differently, our topological invariant is directly defined by using the center-of-mass (c.o.m) quasi-momentum associated with the cotranslational symmetry.
We believe that our definition opens a new route to the characterization of strongly interacting topological states.

\section{Topological invariant associated with cotranslational symmetry}

To illustrate our idea, we first consider a generally two-dimensional (2D) quantum system with $N$ interacting particles.
The Hamiltonian reads as,
\begin{equation}
  H=\sum\nolimits_{j=1}^{N} H_j+\sum\nolimits_{j=1}^{N-1} \sum\nolimits_{j'=j+1}^{N} V(|\mbr_j-\mbr_{j'}|).
\end{equation}
Here, $\mbr_j=(x_j,y_j)$ is the position of the $j$-th particle, the single-particle Hamiltonian $H_j$ is of translational symmetry with respect to the period $\mma=(a_x,a_y)$, and the interaction $V(|\mbr_j-\mbr_{j'}|)$ only depends on the inter-particle distance.
Typical examples are quantum lattice models such as Hubbard lattices and quantum spin lattices.
Although we concentrate on quantum lattice models, our idea can be extended to continuous models.

Given an $N$-particle wave-function $\psi(\mbr_1,\mbr_2,\dots,\mbr_N)$, the single-particle translation operator for the $j$-th particle, $T^{(j)}$, is defined as $T^{(j)}\psi(\mbr_1,\dots,\mbr_j,\dots,\mbr_N)=\psi(\mbr_1,\dots,\mbr_j+\mma,\dots,\mbr_N)$ with $j\in \{1,2,\dots,N\}$.
For a noninteracting system, because of the translational symmetry of each single-particle Hamiltonian $H_j$, $T^{(j)}$ commutes with the whole Hamiltonian and the many-body eigenstate has a tensor product structure of $N$ single-particle Bloch states.
Therefore the independent Bloch momenta of the $N$ particles form a set of good quantum numbers for the noninteracting Hamiltonian.
However, the interaction will break the single-particle translational symmetry and make the $N$ independent Bloch momenta no longer good quantum numbers.

We now define the cotranslation operator, $T_\mma(\tau)$, as
\begin{equation}
  T_\mma(\tau)\psi(\mbr_1,\mbr_2,\dots,\mbr_N)=\psi(\mbr_1+\tau\mma,\mbr_2+\tau\mma,\dots,\mbr_N+\tau\mma)
\end{equation}
with $\tau$ an arbitrary integer.
Actually, $T_\mma(\tau)$ is a combination of all single-particle translation operators, $T_\mma(\tau)=[T^{(1)}T^{(2)}\cdots T^{(N)}]^\tau$, and thus it commutes with each $H_j$.
Since $T_\mma(\tau)T_\mma(\tau')=T_\mma(\tau')T_\mma(\tau)=T_\mma(\tau+\tau')$
and $[T_\mma(\tau)]^{-1}=T_\mma(-\tau)$,
the set $\left\{T_\mma(\tau),\,\tau\in\mathbb{Z}\right\}$ forms an Abelian group (where $\mathbb{Z}$ is the set of all integers).
We call this group as the \emph{cotranslation group}.
As all cotranslation operators commute with the interaction term,
the whole Hamiltonian is invariant under the cotranslation transform,
\begin{equation}
  [T_\mma(\tau)]^{-1}HT_\mma(\tau)=H,
\end{equation}
which represents the \emph{cotranslational symmetry}.

Under the cotranslational symmetry, the Hamiltonian $H$ and $T_\mma(\tau)$ share a set of common eigenstates.
The common eigenstates obey
\begin{equation}
  T_\mma(\tau)\psi(\mbr_1,\mbr_2,\dots,\mbr_N)=c_\mma(\tau)\psi(\mbr_1,\mbr_2,\dots,\mbr_N),
\end{equation}
with $c_\mma(\tau)$ being an eigenvalue of $T_\mma(\tau)$.
It is easy to find $c_\mma(\tau)c_\mma(\tau')=c_\mma(\tau+\tau')$ and $[c_\mma(\tau)]^{-1}=c_\mma(-\tau)$.
Thus the eigenvalues could be chosen as the exponential form $c_\mma(\tau)=e^{i\mathbf{k}\cdot\tau\mma}$ with the vector $\mathbf{k}=(k_x,k_y)$~\cite{Kannappan2009},
which is a pair of good quantum numbers.
Thus we have,
\begin{equation}\label{Eq_Bloch_like_theorem}
  \psi(\mbr_1+\tau\mma,\dots,\mbr_N+\tau\mma)=e^{i\mathbf k\cdot\tau\mma}\psi(\mbr_1,\dots,\mbr_N),
\end{equation}
which resembles the Bloch theorem for single-particle systems with translational symmetry.
Therefore, the vector $\mathbf k$ acts as the corresponding c.o.m quasi-momentum.
Similar to the Bloch functions for single-particle systems with translational symmetry,
one can define $\psi(\mbr_1,\mbr_2,\dots,\mbr_N)=e^{i\mathbf k\cdot\frac{1}{N}(\mbr_1+\mbr_2+\cdots+\mbr_N)}\phi(\mbr_1,\mbr_2,\dots,\mbr_N)$ and then obtain $\phi(\mbr_1+\tau\mma,\mbr_2+\tau\mma,\dots,\mbr_N+\tau\mma)=\phi(\mbr_1,\mbr_2,\dots,\mbr_N)$ from equation~(\ref{Eq_Bloch_like_theorem}).
We thus identify these eigenstates $\psi(\mbr_1,\mbr_2,\dots,\mbr_N)$ as the many-body Bloch states for IQS's with cotranslational symmetry.

By exploiting the cotranslational symmetry and the many-body Bloch states,
we define a topological invariant (the first Chern number).
It is an integral of the Berry curvature $\mathcal{F}_n(k_x,k_y)$ over the first Brillouin zone (BZ),
\begin{equation}\label{Eq_Chern_num}
  C_n=\frac{1}{2\pi}\int\!\!\int_{\mathrm{BZ}}\dif^2\mathbf{k}\,\mathcal{F}_n(k_x,k_y),
\end{equation}
where, $\mathcal{F}_n(k_x,k_y)=\mathrm{Im}\left(
\braket{\partial_{k_x}\phi_n}{\partial_{k_y}\phi_n}-
\braket{\partial_{k_y}\phi_n}{\partial_{k_x}\phi_n}\right)$ is determined by the Bloch state $\ket{\phi_n}=\ket{\phi_n(k_x,k_y)}$,
$k_x\in(-\pi/a_x,\pi/a_x]$, $k_y\in(-\pi/a_y,\pi/a_y]$, and $n$ is the band index.
In fact, the above Chern number is a TKNN-type topological invariant.
We should remark that our topological invariant is always well-defined for the band which is well-separated from other bands, that is, it is protected by the corresponding energy gaps.

\section{Topological bound-states in generalized Heisenberg XXZ model}

\subsection{A generalized Heisenberg XXZ model}

We now consider a generalized 2D Heisenberg XXZ model described by the following Hamiltonian,
\begin{eqnarray}\label{Eq.Heisenberg.model}
  \hat{H}_\mathrm{H}&=&-J_x\sum_{l,m}\left[\left(e^{i2m\Phi}\hat{S}^+_{l+1,m}\hat{S}^-_{l,m}
  +\lambda\hat{S}^+_{l,m+1}\hat{S}^-_{l,m}\right)+\mathrm{h.c.}\right] \nonumber \\
  &&-V_x\sum_{l,m}\left[\hat{S}^z_{l,m}\hat{S}^z_{l+1,m}
  +\lambda\hat{S}^z_{l,m}\hat{S}^z_{l,m+1}\right]
\end{eqnarray}
with the spin-$1/2$ operators $(\hat{S}^x_{l,m},\hat{S}^y_{l,m},\hat{S}^z_{l,m})$ and $\hat{S}^\pm_{l,m}=\hat{S}^x_{l,m}\pm i\hat{S}^y_{l,m}$ for the lattice site $(l,m)$.
Here, $J_x$ and $V_x$ are the transverse and longitudinal spin-exchange couplings, respectively.
And $\lambda$ represents the ratio of the interactions between $y$- and $x$-directions.
Different from the usual 2D Heisenberg XXZ model, our $\hat{H}_\mathrm{H}$ includes a spatially varying phase $2m\Phi$ along $x$-direction.

According to the Matsubara-Matsuda mapping~\cite{Matsubara1956}, the model $\hat{H}_\mathrm{H}$ is equivalent to a hard-core Bose-Hubbard model.
By introducing $\ket{\downarrow}\leftrightarrow\ket{0}$,
$\ket{\uparrow}\leftrightarrow\ket{1}$,
$\hat{S}^+_{l,m}\leftrightarrow\hatd{b}_{l,m}$,
$\hat{S}^-_{l,m}\leftrightarrow\hat{b}_{l,m}$,
and $\hat{S}^z_{l,m}\leftrightarrow\hatd{b}_{l,m}\hat{b}_{l,m}-\frac{1}{2}$,
we have,
\begin{eqnarray}\label{Eq_Hamiltonian}
  \hat{H}&=&-J_x\sum_{l,m}\left(e^{i2\pi\beta m}\hatd{b}_{l+1,m}\hat{b}_{l,m}+\lambda\hatd{b}_{l,m+1}\hat{b}_{l,m}+\mathrm{h.c.}\right) \nonumber \\
  &&-V_x\sum_{l,m}\Big[\hat{n}_{l,m}\hat{n}_{l+1,m}+\lambda\hat{n}_{l,m}\hat{n}_{l,m+1}\Big]
\end{eqnarray}
with the hard-core bosonic creation (annihilation) operators $\hatd{b}_{l,m}$ ($\hat{b}_{l,m}$) and the number operator $\hat{n}_{l,m}=\hatd{b}_{l,m}\hat{b}_{l,m}$.
Here, $\beta=\Phi/\pi$ and we have removed a constant energy shift.
Below, we concentrate on discussing the rational flux $\beta=p/q$ (where $p$ and $q$ are coprime integers) and consider a lattice of $L_x \times L_y$ sites and $L_y=qs$ with an odd integer $s$.

\subsection{Topological two-magnon excitations}

The two-particle Hilbert subspace is spanned by the basis,
$\mathcal{B}^{(2)}_{2\mathrm{D}}=
  \bigl\{\ket{l_1,m_1;l_2,m_2}=
  \hatd{b}_{l_1,m_1}\hatd{b}_{l_2,m_2}\ket{\mathbf{0}}\bigr\}$,
with ($1\le l_1<l_2\le L_x$) or ($1\le l_1=l_2\le L_x$ and $1\le m_1<m_2\le L_y$).
We then impose the periodic boundary conditions (PBCs) in both $x$- and $y$-directions.
By introducing
$\psi_{l_1,m_1;l_2,m_2}=\bra{\mathbf{0}}
\hat{b}_{l_2,m_2}\hat{b}_{l_1,m_1}\ket{\Psi}$,
the eigenstates can be expanded as
$
  \ket{\Psi}=\sum\nolimits_{l_1,m_1;l_2,m_2}
  {\psi_{l_1,m_1;l_2,m_2}\ket{l_1,m_1;l_2,m_2}}.
$
The eigenequation $\hat{H}\ket{\Psi}=E\ket{\Psi}$ gives
\begin{eqnarray}
  E\psi_{l_1,m_1;l_2,m_2}&=&-V_x\big(\delta^{l_1\pm1,m_1}_{l_2,m_2} +\lambda\delta^{l_1,m_1\pm1}_{l_2,m_2}\big)\psi_{l_1,m_1;l_2,m_2} \nonumber \\
  &&-J_x\big(e^{i2\pi\beta m_1}\psi_{l_1-1,m_1;l_2,m_2}
  +e^{-i2\pi\beta m_1}\psi_{l_1+1,m_1;l_2,m_2} \nonumber \\
  &&
  +e^{i2\pi\beta m_2}\psi_{l_1,m_1;l_2-1,m_2}
  +e^{-i2\pi\beta m_2}\psi_{l_1,m_1;l_2+1,m_2} \nonumber \\
  &&
  +\lambda\psi_{l_1,m_1-1;l_2,m_2}
  +\lambda\psi_{l_1,m_1+1;l_2,m_2} \nonumber \\
  &&
  +\lambda\psi_{l_1,m_1;l_2,m_2-1}
  +\lambda\psi_{l_1,m_1;l_2,m_2+1}\big),
\end{eqnarray}
where the PBCs require
$\psi_{l_1+L_x,m_1;l_2,m_2}
=\psi_{l_1,m_1;l_2+L_x,m_2}
=\psi_{l_1,m_1+L_y;l_2,m_2}
=\psi_{l_1,m_1;l_2,m_2+L_y}
=\psi_{l_1,m_1;l_2,m_2}$,
and the hard-core bosonic commutation relations require $\psi_{l_1,m_1;l_2,m_2}=\psi_{l_2,m_2;l_1,m_1}$
and $\psi_{l_1,m_1;l_1,m_1}=0$.

To describe the cotranslational symmetry along $x$-direction,
we introduce the two-particle cotranslational operator $T_1^x$ as
\begin{equation}
  T_1^x\psi_{l_1,m_1;l_2,m_2}=\psi_{l_1+1,m_1;l_2+1,m_2}.
\end{equation}
It's easy to find that
$HT_1^x\psi_{l_1,m_1;l_2,m_2}=T_1^xH\psi_{l_1,m_1;l_2,m_2}$ holds for arbitrary $\psi_{l_1,m_1;l_2,m_2}$.
Therefore, the Hamiltonian $H$ commutes with $T_1^x$
and they share a set of common eigenstates:
$\psi_{l_1,m_1;l_2,m_2}=e^{\frac{i}{2}k_x(l_1+l_2)}\phi_{l_1,m_1;l_2,m_2}$,
in which $\phi_{l_1+1,m_1;l_2+1,m_2}=\phi_{l_1,m_1;l_2,m_2}$ is invariant under $T_1^x$.
The eigenequation $H_{k_x}\ket{\Phi}=E_{k_x}\ket{\Phi}$ gives
\begin{eqnarray}
  E_{k_x}\phi_{l_1,m_1;l_2,m_2}&=&
  -V_x(\delta^{l_1\pm1,m_1}_{l_2,m_2}+\lambda\delta^{l_1,m_1\pm1}_{l_2,m_2})
  \phi_{l_1,m_1;l_2,m_2}
  \nonumber \\ &&
  -J_x\big(e^{i2\pi\beta m_1-i\frac{k_x}{2}}\phi_{l_1-1,m_1;l_2,m_2}
  +e^{-i2\pi\beta m_1+i\frac{k_x}{2}}\phi_{l_1+1,m_1;l_2,m_2}
  \nonumber \\ &&
  +e^{i2\pi\beta m_2-i\frac{k_x}{2}}\phi_{l_1,m_1;l_2-1,m_2}
  +e^{-i2\pi\beta m_2+i\frac{k_x}{2}}\phi_{l_1,m_1;l_2+1,m_2}
  \nonumber \\ &&
  +\lambda\phi_{l_1,m_1-1;l_2,m_2}+\lambda\phi_{l_1,m_1+1;l_2,m_2}
  \nonumber \\ &&
  +\lambda\phi_{l_1,m_1;l_2,m_2-1} +\lambda\phi_{l_1,m_1;l_2,m_2+1}\big).
\end{eqnarray}
Here, $H_{k_x}$ denotes the $k_x$-block of the two-particle Hamiltonian and $k_x=\frac{2\pi}{L_x}\alpha_x$ is the c.o.m quasi-momentum along $x$-direction (with the integer $\alpha_x\in [-\frac{L_x-1}{2},\frac{L_x-1}{2}]$).
Correspondingly, the PBCs require $\phi_{l_1+L_x,m_1;l_2,m_2}=\phi_{l_1,m_1;l_2+L_x,m_2}
=(-1)^{\alpha_x}\phi_{l_1,m_1;l_2,m_2}$
and $\phi_{l_1,m_1+L_y;l_2,m_2}
=\phi_{l_1,m_1;l_2,m_2+L_y}=\phi_{l_1,m_1;l_2,m_2}$,
and the commutation relations require $\phi_{l_1,m_1;l_2,m_2}=\phi_{l_2,m_2;l_1,m_1}$
and $\phi_{l_1,m_1;l_1,m_1}=0$.

Similarly, to describe the cotranslational symmetry along y-direction, we introduce $T^{y}_q$ as
\begin{equation}
  T^{y}_q\psi_{l_1,m_1;l_2,m_2}=\psi_{l_1,m_1+q;l_2,m_2+q}.
\end{equation}
In the $k_x$-subspace, it turns out to be
$T^{y}_q\phi_{l_1,m_1;l_2,m_2}=\phi_{l_1,m_1+q;l_2,m_2+q}$.
It is easy to find that
$H_{k_x}T^{y}_q\phi_{l_1,m_1;l_2,m_2}=T^{y}_qH_{k_x}\phi_{l_1,m_1;l_2,m_2}$
holds for arbitrary $\phi_{l_1,m_1;l_2,m_2}$.
Therefore, $H_{k_x}$ and $T^{y}_q$ have a common set of eigenstates,
which can be written as
$\phi_{l_1,m_1;l_2,m_2}=e^{\frac{i}{2}k_y(m_1+m_2)}\varphi_{l_1,m_1;l_2,m_2}$,
where $\varphi_{l_1,m_1+q;l_2,m_2+q}=\varphi_{l_1,m_1;l_2,m_2}$ is invariant under $T^{y}_q$.
The eigenequation $H_{k_x,k_y}\ket{\varphi}=E_{k_x,k_y}\ket{\varphi}$ reads
\begin{eqnarray}\label{Eq.Ham.kx.ky.S}
  &&E_{k_x,k_y}\varphi_{l_1,m_1;l_2,m_2}
  =-V_x(\delta^{l_1\pm1,m_1}_{l_2,m_2}+\lambda\delta^{l_1,m_1\pm1}_{l_2,m_2})
  \varphi_{l_1,m_1;l_2,m_2} \nonumber \\
  &&-J_x\bigl(e^{i2\pi\beta m_1-\frac{i}{2}k_x}\varphi_{l_1-1,m_1;l_2,m_2}
  +e^{-i2\pi\beta m_1+\frac{i}{2}k_x}\varphi_{l_1+1,m_1;l_2,m_2} \nonumber \\
  &&+e^{i2\pi\beta m_2-\frac{i}{2}k_x}\varphi_{l_1,m_1;l_2-1,m_2}
  +e^{-i2\pi\beta m_2+\frac{i}{2}k_x}\varphi_{l_1,m_1;l_2+1,m_2} \nonumber \\
  &&+\lambda e^{-\frac{i}{2}k_y}\varphi_{l_1,m_1-1;l_2,m_2}+\lambda e^{\frac{i}{2}k_y}\varphi_{l_1,m_1+1;l_2,m_2} \nonumber \\
  &&+\lambda e^{-\frac{i}{2}k_y}\varphi_{l_1,m_1;l_2,m_2-1}+\lambda e^{\frac{i}{2}k_y}\varphi_{l_1,m_1;l_2,m_2+1}\bigr).
\end{eqnarray}
Here, $k_y=\frac{2\pi}{L_y}\alpha_y$ is the c.o.m quasi-momentum along $y$-direction (with the integer $\alpha_y\in [-\frac{s-1}{2},\frac{s-1}{2}]$).

\begin{figure}[t]
  \includegraphics[width=1.0\columnwidth]{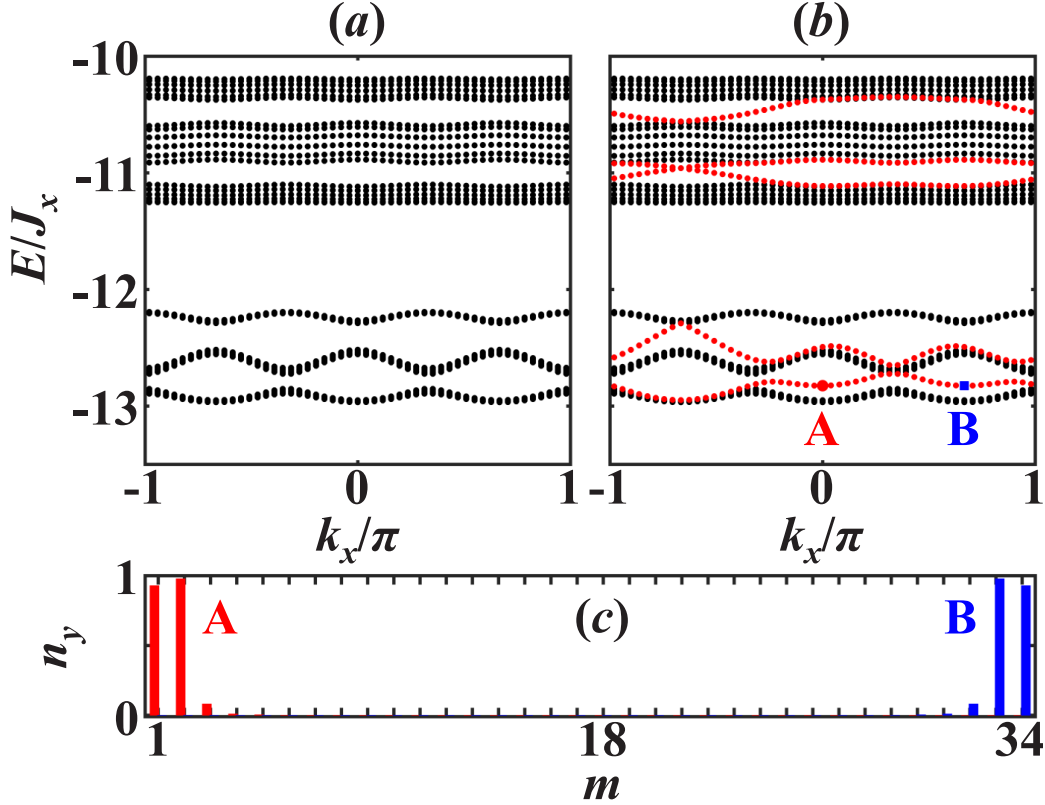}
  \caption{\label{Fig1}
  Two-magnon bound-state spectra.
  Bound-state bands for (a) periodic BC along $y$-direction with $L_y=33$ and (b) open BC along $y$-direction with $L_y=34$.
  Both (a) and (b) choose the periodic BC along $x$-direction with $L_x=51$.
  (c) The density distribution along $y$-direction for $A$ (red circle) with $k_x=0$ and $B$ (blue square) with $k_x=2\pi/3$ in (b).
  The other parameters are chosen as $\beta=1/3$, $V_x/J_x=10$ and $\lambda=1.2$.}
\end{figure}

We now discuss the energy spectrum.
Under strong interactions, in addition to the continuum band, there appear BS bands.
In Fig.~\ref{Fig1}(a), we show the BS spectrum under periodic BCs.
It includes $6$ subbands for $\beta=1/3$ (in general, it includes $2q$ subbands for $\beta=p/q$).
Based on our calculation, the Chern numbers for the $6$ subbands are $(C_1,C_2,\dots,C_6)=(-1,2,-1,-1,2,-1)$.
The Chern numbers indicate the bulk system has a nontrivial topology.
According to the bulk-edge correspondence, topological edge states will appear in the system under open BC.
So we calculate the spectrum under the open BC along $y$-direction.
In Fig.~\ref{Fig1}(b), in addition to the extended BS's, topological edge BS's do appear.
In Fig.~\ref{Fig1}(c), corresponding to the two points $(A,~B)$ in Fig.~\ref{Fig1}(b), we show their density distributions along $y$-direction [$n_y(m)=\left<\hat{n}_y(m)\right>=\left<\sum_{l}\hat{n}_{l,m}\right>$].
The density distributions clearly show that these BS's do localize on the edges.

\subsection{Topological phase transitions (TPTs)}

\begin{figure}[t]
  \includegraphics[width=1.0\columnwidth]{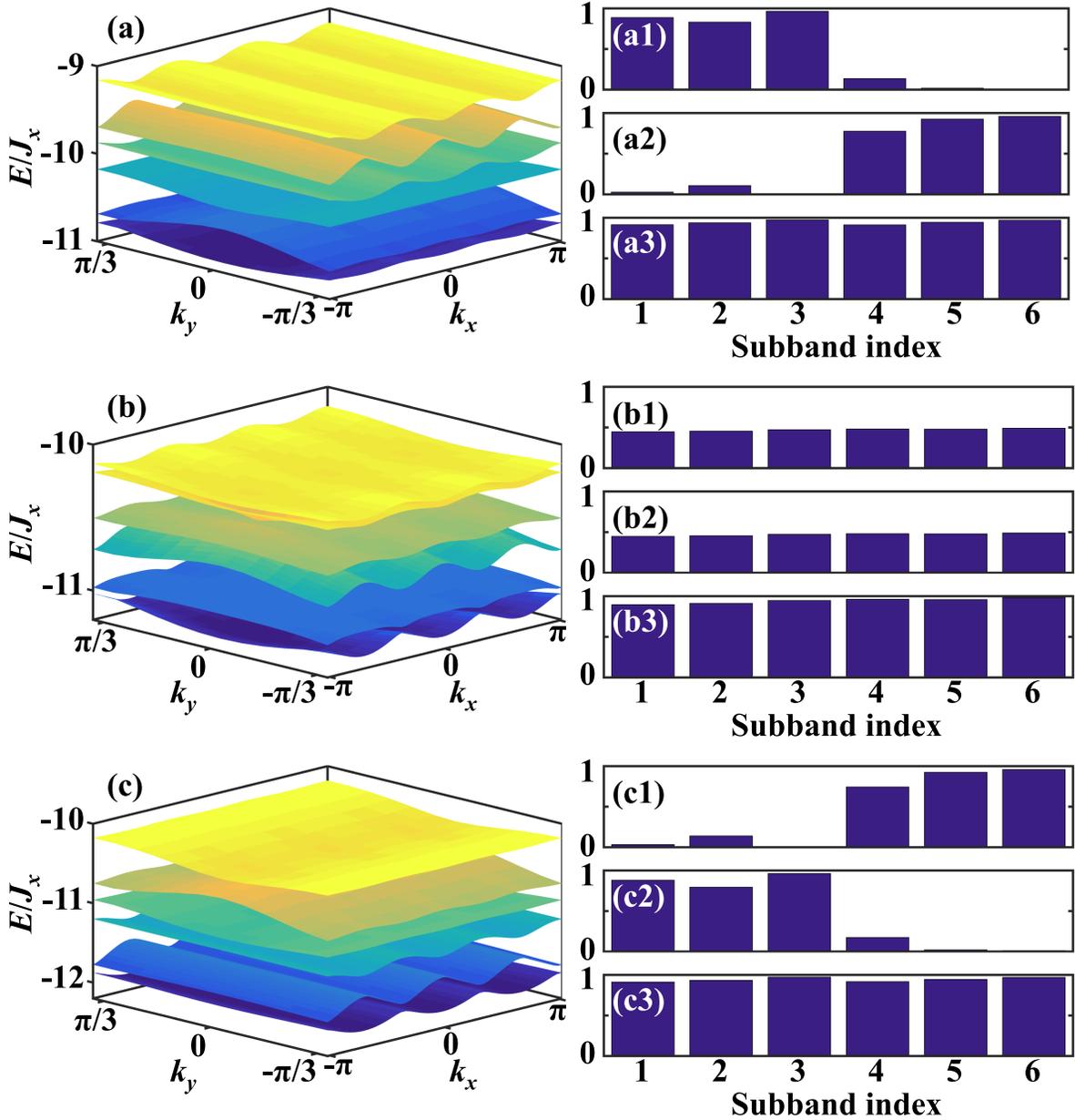}
  \caption{\label{Fig2}
  Two-magnon Bloch bands (i.e. two-magnon bound-state subbands) for (a) $\lambda=0.9$, (b) $\lambda=1$, and (c) $\lambda=1.1$. Signature of bound-states for the six eigen bound-states with $(k_x,k_y)=(0,0)$: (a1,~b1,~c1) $P_x$, (a2,~b2,~c2) $P_y$, and (a3,~b3,~c3) $P=P_x+P_y$. The other parameters are chosen as $\beta=1/3$, $V_x/J_x=10$, $L_x=51$, and $L_y=33$.}
\end{figure}

The interaction ratio $\lambda$ plays an important role in the BS spectrum.
If $\lambda\gg1$ (i.e. the interaction along $y$-direction dominates), the eigenstates of the three lowest subbands can be approximated by a superposition of $\ket{l,m;l,m+1}$, which are called $y$-type BS's.
While the eigenstates of the three higher subbands can be approximated by a superposition of $\ket{l,m;l+1,m}$, which are called $x$-type BS's.
If $\lambda\approx1$ (i.e. $V_x \approx V_y$), the BS's are approximated by superpositions of $\ket{l,m;l+1,m}$ and $\ket{l,m;l,m+1}$.
Otherwise, if $\lambda\ll1$, the three lowest subbands correspond to $x$-type BS's while three higher subbands correspond to $y$-type BS's.
By introducing $P_x=\sum_{l,m}|\psi_{l,m;l+1,m}|^2$ and $P_y=\sum_{l,m}|\psi_{l,m;l,m+1}|^2$, we have $P_x=1$ ($P_y=1$) for a perfect $x$-type ($y$-type) BS.
For an arbitrary BS, we find that $P=P_x+P_y \simeq 1$ (see Fig.~\ref{Fig2}).

\begin{figure}[t]
  \includegraphics[width=1.0\columnwidth]{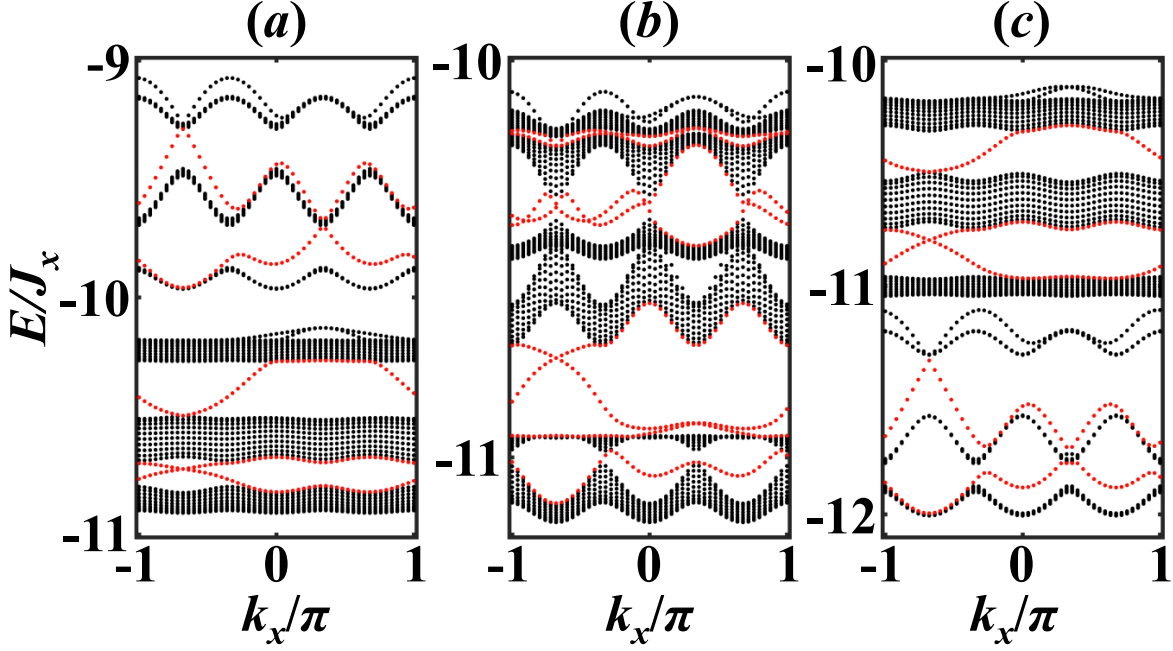}
  \caption{\label{Fig3}
  Bound-state spectra with open BC along $y$ direction for (a) $\lambda=0.9$, (b) $\lambda=1$ and (c) $\lambda=1.1$. The other parameters are chosen as $\beta=1/3$, $V_x/J_x=10$, $L_x=51$, and $L_y=34$.}
\end{figure}

In Fig.~\ref{Fig3}, we show the BS spectra for different $\lambda$.
Given the Riemann surface of Bloch states, the energy gaps represent the holes in the Riemann surface and the winding number of the edge states around these holes is another topological invariant~\cite{Hatsugai1993PRL,Hatsugai1993PRB}.
We find the absolute value of the winding number $W_1$ for the edge states in the first energy gap: $|W_1|=1$ for $\lambda = 0.9$ and $1.1$ [see Fig.~\ref{Fig3}(a,c)] and $|W_1|=2$ for $\lambda=1$ [see Fig.~\ref{Fig3}(b)].
This means that TPTs appear in the two regions: $0.9<\lambda<1$ and $1<\lambda<1.1$.
Our calculations show that the Chern numbers for the lowest subband are $C_1=(-1,2,-1)$ for $\lambda=(0.9,1,1.1)$, which are consistent with the winding numbers for the corresponding edge states.

\begin{figure}[t]
  \includegraphics[width=1.0\columnwidth]{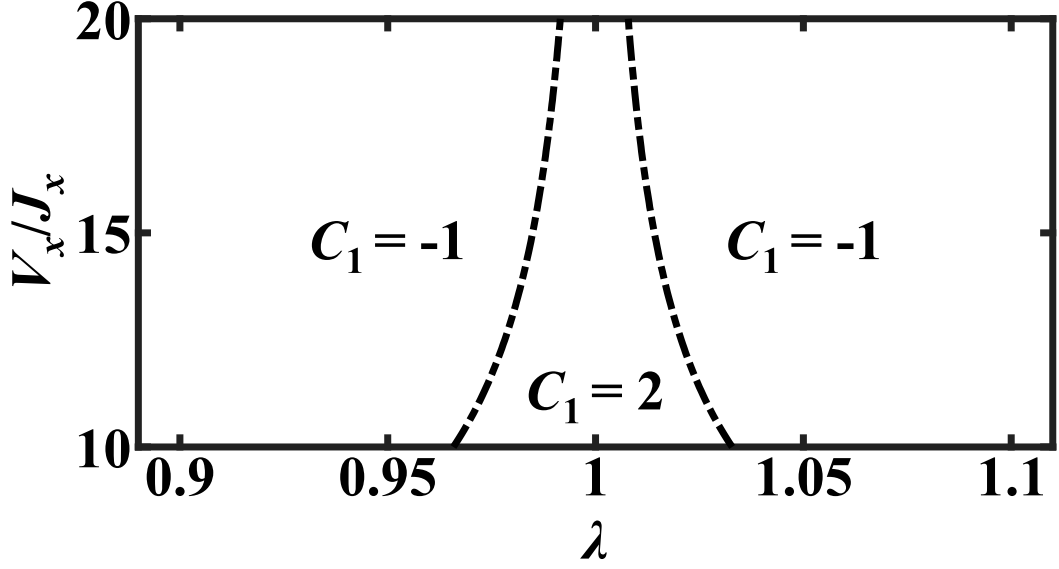}
  \caption{\label{Fig4} Topological phase diagram for the lowest bound-state subband under periodic BCs with $\beta=1/3$.}
\end{figure}

According to the topological band theory~\cite{KaneRev2010,QiRev2011}, TPTs associate with gap closures.
For a finite system, a gap closure corresponds to a gap minimum which approaches to zero when the system size increases.
In Fig.~\ref{Fig4}, we show the topological phase diagram for the first BS subband.

\subsection{Effective single-particle Hofstadter superlattices}

\begin{figure}[t]
  \includegraphics[width=1.0\columnwidth]{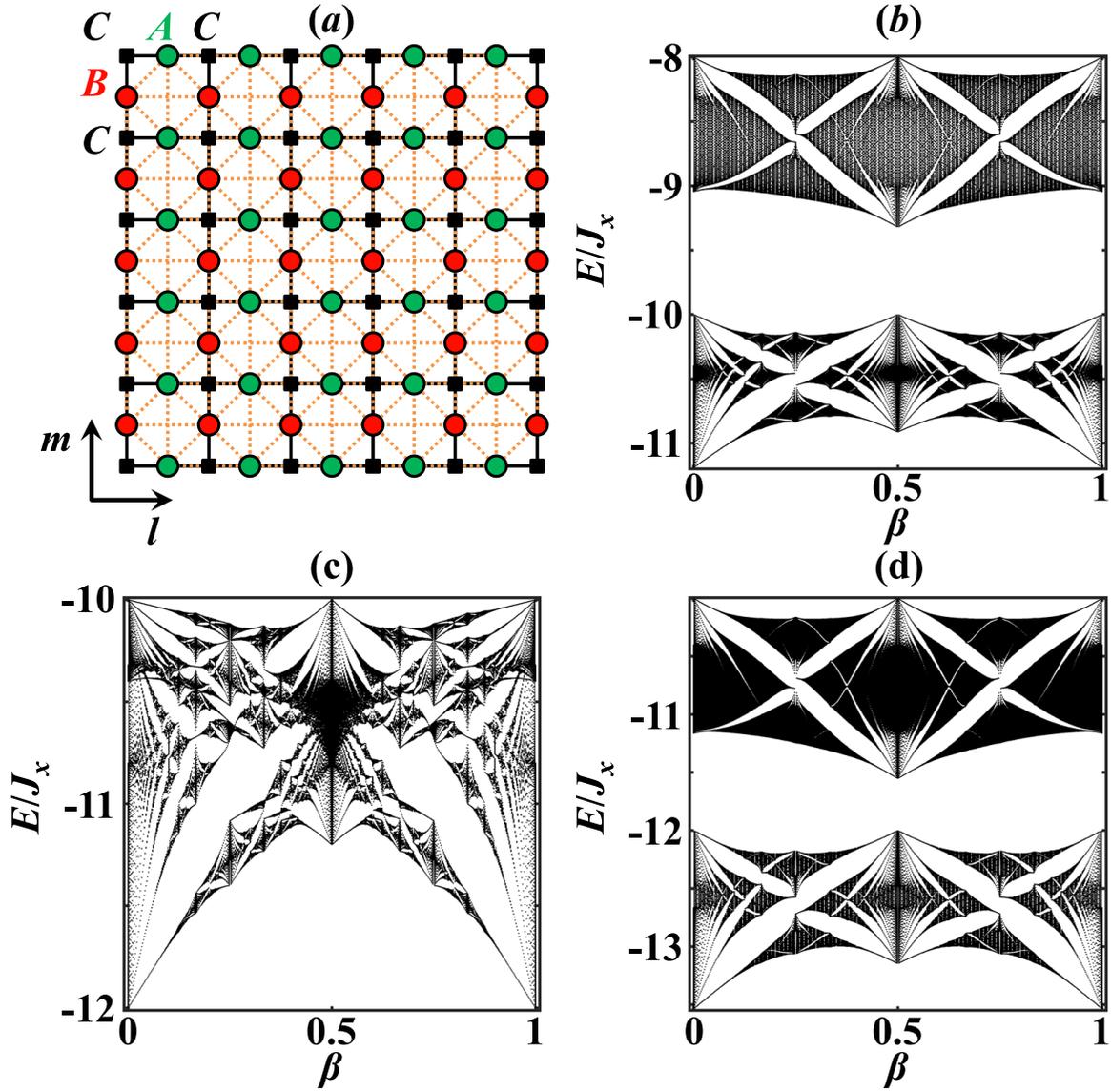}
  \caption{\label{Fig5}
  Hofstadter superlattice and its butterfly-like spectra.
  (a) The sublattices A and B are respectively denoted by green and red circles. The original lattice C is represented by black squares.
  The sublattice site $(l,m)_A$ [$(l,m)_B$] locates at the middle point of $(l,m)_C$ and $(l+1,m)_C$ [$(l,m)_C$ and $(l,m+1)_C$].
  The orange dot lines represent the couplings between the two sublattices.
  The butterfly-like spectra are shown in (b) for $\lambda=0.9$, (c) for $\lambda=1$, and (d) for $\lambda=1.1$.
  The other parameters are chosen as $V_x/J_x=10$, $k_x=0$, and $L_y=1000$.}
\end{figure}

Under strong interactions ($|J_x/V_x|\ll1$), a BS can be regarded as a quasiparticle.
By treating the hopping as a perturbation to the interaction and implementing the Schrieffer-Wolff transformation~\cite{LossSWRev2011AoP}, the system obeys an effective single-particle model (see Appendix C),
\begin{eqnarray}\label{Eq_Effective_Ham}
  \hat{H}_\mathrm{eff}&=&\hat{H}_\mA+\hat{H}_\mB+\hat{H}_{\mA\mB}, \\
  \hat{H}_\mA&=&-J_\mathrm{eff}\sum_{l,m}\Big[\Big(e^{i4\pi\beta m}\hatd{A}_{l+1,m}\hat{A}_{l,m}+\mathrm{h.c.}\Big)\nonumber \\
  &&~~~~~~~~~+2\lambda^2 \Big(\hatd{A}_{l,m+1}\hat{A}_{l,m} +\mathrm{h.c.}\Big)+\epsilon_x\hatd{A}_{l,m}\hat{A}_{l,m}\Big],\nonumber \\
  \hat{H}_\mB&=&-J_\mathrm{eff}\sum_{l,m}\Big[\frac{2}{\lambda}\Big(e^{i4\pi\beta m}e^{i2\pi\beta}\hatd{B}_{l+1,m}\hat{B}_{l,m}+\mathrm{h.c.}\Big)\nonumber \\
  &&~~~~~~~~~+\lambda\Big(\hatd{B}_{l,m+1}\hat{B}_{l,m}+\mathrm{h.c.}\Big) +\epsilon_y\hatd{B}_{l,m}\hat{B}_{l,m}\Big],\nonumber \\
  \hat{H}_{\mA\mB}&=&-J_\mathrm{eff}J_{xy}\sum_{l,m}\Big[e^{i2\pi\beta m}e^{i\pi\beta} \Big(\hatd{A}_{l,m}\hat{B}_{l,m} +\hatd{B}_{l+1,m}\hat{A}_{l,m}\nonumber \\
  &&~~~~~~~~~+\hatd{A}_{l,m+1}\hat{B}_{l,m} +\hatd{B}_{l+1,m}\hat{A}_{l,m+1}\Big)+\mathrm{h.c.}\Big].\nonumber
\end{eqnarray}
Here, $J_\mathrm{eff}=J_x^2/V_x$, $J_{xy}=(\lambda+1)\cos(\pi\beta)$,
$\epsilon_x=V_x^2/J_x^2+2+4\lambda^2$,
and $\epsilon_y=\lambda V_x^2/J_x^2+2\lambda+4/\lambda$.
The operators $\hatd{A}_{l,m}$ and $\hatd{B}_{l,m}$ create a particle in states $\ket{l,m;l+1,m}$ and $\ket{l,m;l,m+1}$, respectively.
In Fig.~\ref{Fig5}(a), we show the lattice structure, in which the green and red circles respectively represent the sublattice-A and B.
Actually, $\hat{H}_\mA$ and $\hat{H}_\mB$ are standard Hofstadter Hamiltonians, and $\hat{H}_{\mA\mB}$ describes the coupling between the two sublattices.

Now we discuss the quasi-particle spectrum.
Under the periodic BC along $x$-direction, through the Fourier transformation:
$\hatd{A}_{l,m}=\frac{1}{\sqrt{L_x}}\sum_{k_x}e^{-ik_xl}\hatd{A}_{k_x,m}$
and $\hatd{B}_{l,m}=\frac{1}{\sqrt{L_x}}\sum_{k_x}e^{-ik_xl}\hatd{B}_{k_x,m}$,
the system~(\ref{Eq_Effective_Ham}) becomes block diagonalized.
The eigenstates $\ket{\Psi(k_x)}=\big[\psi^\mA_m(k_x)\hatd{A}_{k_x,m}
+\psi^\mB_m(k_x)\hatd{B}_{k_x,m}\big]\ket{\mathbf{0}}$ obey the coupled Harper equations,
\begin{eqnarray}\label{Eq_Coupled_Harper}
  E'\psi_{m}&=&-\left[\begin{array}{cc}
  J^\mA & J_{m-1}e^{i\frac{k_x}{2}} \\
  0 & J^\mB
  \end{array}\right]\psi_{m-1}
  -\left[\begin{array}{cc}
  J^\mA & 0 \\
  J_{m}e^{-i\frac{k_x}{2}} & J^\mB
  \end{array}\right]\psi_{m+1} \nonumber \\
  &&-\left[\begin{array}{cc}
  \epsilon_m^\mA & J_{m}e^{i\frac{k_x}{2}} \\
  J_{m}e^{-i\frac{k_x}{2}} & \epsilon_m^\mB
  \end{array}\right]\psi_{m},
\end{eqnarray}
with $E'=\slfrac{E}{J_\mathrm{eff}}$,
$\psi_m=[\psi^\mA_m,\psi^\mB_m]^\mathrm{T}$,
$J^\mA=2\lambda^2$, $J^\mB=\lambda$,
$\epsilon_m^\mA=\epsilon_x+2\cos(4\pi\beta m-k_x)$,
$\epsilon_m^\mB=\epsilon_y+(\slfrac{4}{\lambda})\cos(4\pi\beta m+2\pi\beta-k_x)$,
and $J_m=2J_{xy}\cos(2\pi\beta m+\pi\beta-\slfrac{k_x}{2})$.
In Fig.~\ref{Fig5}~(b, c, d), we show the spectrum versus $\beta$.
At $\lambda=0.8$ and $1.2$, the butterfly-like spectrum includes two separated parts.
When $\lambda \rightarrow 1$, the gap between the two parts gradually vanishes.
Finally, at $\lambda=1$, the two parts merge into one butterfly.
Actually, such a spectrum deformation can be induced by tuning the hoping ratio $J/K$ of the spinor Hofstadter model~(\ref{Eq_Bose-Hubbard}).

\subsection{Experimental possibility}

In this section, we briefly discuss the experimental possibility.
Using the laser-assisted tunneling of two-component Bose atoms in a tilted optical lattice, one can realize a 2D interacting spinor Hofstadter model~\cite{Bloch2013,Ketterle2013}, which is governed by the following Hamiltonian (see Appendix A),
\begin{eqnarray}\label{Eq_Bose-Hubbard}
  \hat{H}_\mathrm{B}&=& -\sum_{l,m,\sigma}\Big[Ke^{i\alpha_\sigma m\Phi}\hatd{a}_{l+1,m,\sigma}\hat{a}_{l,m,\sigma} +\mathrm{h.c.}\Big]
  -\sum_{l,m,\sigma}\Big[J\hatd{a}_{l,m+1,\sigma}\hat{a}_{l,m,\sigma}+\mathrm{h.c.}\Big] \nonumber \\
  &&+\sum_{l,m,\sigma_1,\sigma_2}\frac{1}{2}U_{\sigma_1\sigma_2}\hat{n}_{l,m,\sigma_1}
  \Big(\hat{n}_{l,m,\sigma_2}-\delta_{\sigma_1,\sigma_2}\Big).
\end{eqnarray}
with the lattice index $(l,m)$, the component index $\sigma \in \{\uparrow,\downarrow\}$, the creation (annihilation) operators $\hatd{a}_{l,m,\sigma}$ ($\hat{a}_{l,m,\sigma}$), and the number operator $\hat{n}_{l,m,\sigma}=\hatd{a}_{l,m,\sigma}\hat{a}_{l,m,\sigma}$.
Here, $U_{\sigma_1 \sigma_2}$ is the on-site interaction whose strength can be tuned via Feshbach resonances~\cite{Feshbach-Widera2004,Feshbach-Gross2010}, and $K$ and $J$ are respectively the hopping strengths along $x$ and $y$ directions.
The hopping along $x$-direction involves an additional spin- and spatial-dependent phase $\phi_{\sigma,m}=\alpha_\sigma m\Phi$ with $\alpha_\uparrow=1$ and $\alpha_\downarrow=-1$.

In the strong interaction regime with unit filling, by using the second-order perturbation theory~\cite{Takahashi1977}, one can map the model~(\ref{Eq_Bose-Hubbard}) onto the 2D generalized Heisenberg XXZ model~(\ref{Eq.Heisenberg.model}) (see Appendix B) with the parameters are given as $J_x=\slfrac{2K^2}{U_{\uparrow\downarrow}}$, $J_y=\lambda J_x$,
$V_x=4K^2(\slfrac{1}{U_{\uparrow\uparrow}} +\slfrac{1}{U_{\downarrow\downarrow}}-\slfrac{1}{U_{\uparrow\downarrow}})$,
$V_y=\lambda V_x$, and $\lambda=V_y/V_x=J^2/K^2$.

The selective magnon excitations can be prepared by using a line-shaped laser beam generated with a spatial light modulator~\cite{Bloch-NatPhys-2013,Bloch-Nature-2013}.
The two-magnon bound states can be observed by using the \textit{in situ} correlation measurement~\cite{Bloch-Nature-2013}.
Furthermore, one can explore topological phase transition by varying $\lambda$ and $V_x/J_x$, which are respectively determined by the hopping ratio $J/K$ and the two interaction ratios $(U_{\uparrow\downarrow}/U_{\uparrow\uparrow}, U_{\uparrow\downarrow}/U_{\downarrow\downarrow})$ of the interacting spinor Hofstadter model~(\ref{Eq_Bose-Hubbard}).

It's worth to mention that, the interacting spinor Hofstadter model can be realized by two-component systems of either bosons or fermions.
Our above discussions concentrate on the systems realized by two-component bosons, whose inter-particle interactions are described by three different s-wave scattering lengths, which breaks the time-reversal symmetry.
However, for the systems realized by fermions, it is possible to keep the time-reversal-invariance in the Hofstadter-Hubbard model~\cite{Goldman2010,Cocks2012,Orth2013}.

\section{Conclusion}

In summary, from the cotranslational symmetry (collectively translational invariance), we introduce an intrinsic topological invariant for interacting multi-particle quantum systems.
Our topological invariant is defined as an integral of Berry curvature over the first Brillouin expanded by the c.o.m. quasi-momentum.
Our definition generalizes the well-known TKNN invariant~\cite{TKNN1982} and it always works for the bands well-separated from others.
As an application, we use our topological invariant to study the two-magnon excitations in a generalized 2D Heisenberg XXZ model.
We explore the nontrivial topology of these excitations and demonstrate the emergence of topological edge bound-states.
We further give the topological phase diagram for the lowest bound-state subband.
To understand the topological bound-states, we derive an effective single-particle model described by a Hofstadter superlattice with two coupled standard Hofstadter sublattices.
And we also discuss the possible realization of our model via currently cold-atom experimental techniques.

\section*{Acknowledgments}

This work is supported by the National Natural Science Foundation of China (Grant No. 11374375, 11574405).

\appendix

\section*{Appendix}
\setcounter{section}{0}

\section{Realization of the interacting spinor Hofstadter model}

In this section, we give a detailed derivation of the interacting spinor Hofstadter model.
Based upon the approach for treating noninteracting spinless bosons~\cite{Ketterle2013}, we generalize it to deal with interacting two-component bosons.

\begin{figure}[t]
  \includegraphics[width=1.0\columnwidth]{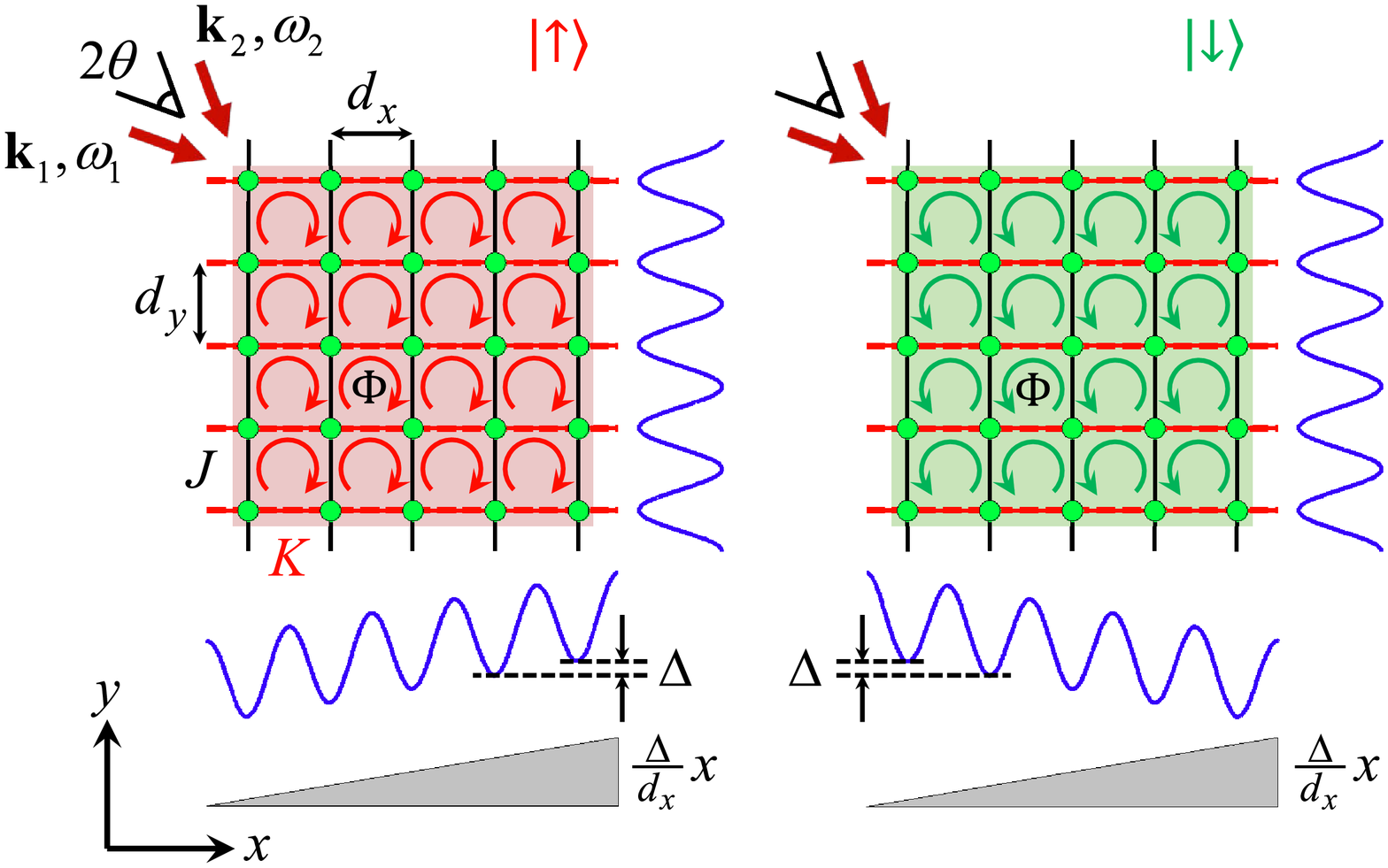}
  \caption{\label{FigA1}Schematic diagram.
  The ultracold Bose atoms are confined in a two-dimensional optical lattice.
  The lattice constants are given as $d_\alpha=\lambda_\alpha/2$ ($\alpha=x$,~$y$).
  Along $y$-direction, the nearest-neighboring tunneling occurs with strength $t_y$.
  Along $x$-direction, the nearest-neighboring tunneling is affected by a magnetic field gradient $\Delta/d_x$, which introduces an energy offset between neighboring sites of (left) $\Delta$ for $\ket{\uparrow}$ atoms and (right) $-\Delta$ for $\ket{\downarrow}$ atoms.
  An additional pair of laser beams with wave vectors $|\mathbf{k}_1|\simeq|\mathbf{k}_2|=2\pi/\lambda_K$ and frequency difference $\omega=\omega_1-\omega_2$ is used to restore resonant tunneling with complex amplitude $K$.
  This realizes an effective flux of $\Phi=k'_y d_y$ [where $\mathbf{k}'=\mathbf{k}_1-\mathbf{k}_2=(k'_x,k'_y)$] for $\ket{\uparrow}$ bosons (left) and $-\Phi$ for $\ket{\downarrow}$ bosons (right).}
\end{figure}

We consider an ultracold two-component Bose gas confined in a 2D optical lattice potential,
\begin{equation}
  V_\mathrm{latt}(\mbr)=
  \frac{V_{x0}}{2}\cos\left(\frac{2\pi}{d_x}x\right)+
  \frac{V_{y0}}{2}\cos\left(\frac{2\pi}{d_y}y\right),
\end{equation}
with $\mbr=(x,y)$ and $d_\alpha=\lambda_\alpha/2$. Here, $\lambda_\alpha$ and $V_{\alpha0}$ are respectively the wavelength and lattice depth along $\alpha$-direction (where $\alpha=x$, $y$).
A gradient magnetic field along $x$-direction is used to generate a spin-dependent linear potential,
\begin{equation}
  V_\mathrm{til}(\mbr)=\frac{\Delta}{d_x}x\hat{\sigma}_z,
\end{equation}
with the amplitude $\Delta$.
Given the bare coupling along $x$-direction $t_x$, when $\Delta\gg t_x$, the tunneling along $x$-direction is inhibited and can be restored by a pair of far-detuned running-wave beams,
\begin{equation}
  V_K(\mbr,t)=\Omega\cos(\mathbf{k}'\cdot\mbr-\omega t),
\end{equation}
with $\omega=\omega_1-\omega_2=\Delta/\hbar$ and $\mathbf{k}'=\mathbf{k}_1-\mathbf{k}_2=(k'_x,k'_y)$, see Fig.~\ref{FigA1}.

If the atom-atom interactions are dominated by two-body interactions, the many-body Hamiltonian includes two parts: a one-body part for single-particle contributions and a two-body part for atom-atom interactions.
The single-particle Hamiltonian reads,
\begin{equation}
  \hat{h}_0=\frac{\hat{\mathbf{p}}^2}{2M}+V_\mathrm{latt}(\mbr)
  +V_\mathrm{til}(\mbr)+V_K(\mbr,t).
\end{equation}
Here, $\hat{\mathbf{p}}=(\hat{p}_x,\hat{p}_y)$ and $M$ is the atomic mass.
Under ultralow temperature, the atom-atom interaction is described by the $s$-wave scattering and the many-body Hamiltonian reads,
\begin{eqnarray}
  \hat{H}&=&\!\!\int\!\dif^2\mbr\left[
  \hatd{\psi}(\mbr)\hat{h}_0\hat{\psi}(\mbr)
  \right]
  +\sum_{\sigma_1,\sigma_2}\frac{1}{2}g_{\sigma_1\sigma_2}
  \!\!\int\!\dif^2\mbr\left[
  \hatd{\psi}_{\sigma_1}(\mbr)\hatd{\psi}_{\sigma_2}(\mbr)
  \hat{\psi}_{\sigma_2}(\mbr)\hat{\psi}_{\sigma_1}(\mbr)
  \right],\nonumber \\
\end{eqnarray}
with the field operators $\hatd{\psi}(\mbr)=[\hatd{\psi}_\uparrow(\mbr),\hatd{\psi}_\downarrow(\mbr)]$,
which creates a boson at position $\mbr$ with $[\ket{\uparrow},\ket{\downarrow}]$.
The interaction strength is given as $g_{\sigma_1\sigma_2}=\frac{4\pi\hbar^2}{M}a_{\sigma_1\sigma_2}$ with $a_{\sigma_1\sigma_2}$ denoting the $s$-wave scattering length between components $\sigma_1$ and $\sigma_2$.
Introducing $\gamma_\uparrow=1$ and $\gamma_\downarrow=-1$,
the many-body Hamiltonian becomes,
\begin{eqnarray}
  \hat{H}&=&\sum_{\sigma}\int\dif^2\mbr\left[
  \hatd{\psi}_\sigma(\mbr)\hat{h}_{0,\sigma}\hat{\psi}_\sigma(\mbr)\right] \nonumber \\
  &&+\sum_{\sigma_1,\sigma_2}\frac{1}{2}g_{\sigma_1\sigma_2}
  \int\dif^2\mbr\left[
  \hatd{\psi}_{\sigma_1}(\mbr)\hatd{\psi}_{\sigma_2}(\mbr)
  \hat{\psi}_{\sigma_2}(\mbr)\hat{\psi}_{\sigma_1}(\mbr)
  \right]
\end{eqnarray}
with
\begin{equation}
  \hat{h}_{0,\sigma}=\frac{\hat{\mathbf{p}}^2}{2M}+V_\mathrm{latt}(\mbr)
  +\gamma_\sigma\frac{\Delta}{d_x}x+V_K(\mbr,t).
\end{equation}
Although the system may involve multiple bands, we assume our system only involves the lowest band which can be realized when the optical lattice is sufficiently deep.

Now we consider the Wannier-Stark-Wannier (WS-W) functions for the lowest band,
\begin{equation}
  \phi_\sigma(\mbr-\mbr_{l,m})=\phi_\sigma^{\mathrm{ws}x}(x-x_l)\phi^{\mathrm{w}y}(y-y_m),
\end{equation}
with $\mbr_{l,m}=(x_l,y_m)=(ld_x,md_y)$.
Define $\hat{h}_\alpha=\frac{\hat{p}_\alpha^2}{2M} +\frac{V_{\alpha0}}{2}\cos(\frac{2\pi}{d_\alpha}\alpha)$ for $\alpha=x$ and $y$, we have $\phi_\sigma^{\mathrm{ws}x}(x-x_l)$ being the Wannier-Stark function for $\hat{h}_{x,\sigma}=\hat{h}_x+\alpha_\sigma\frac{\Delta}{d_x}x$,
while $\phi^{\mathrm{w}y}(y-y_m)$ being the Wannier function for $\hat{h}_{y}$.
By using the Wannier functions $\phi^{\mathrm{w}x}(x-x_j)$ for $\hat{h}_x$, the Wannier-Stark functions $\phi_\sigma^{\mathrm{ws}x}(x-x_l)$ can be expanded as,
\begin{equation}
  \phi_\sigma^{\mathrm{ws}x}(x-x_l)=\sum_jJ_{l-j}(\gamma_\sigma)\phi^{\mathrm{w}x}(x-x_j),
\end{equation}
with $\gamma_\sigma=\alpha_\sigma2t_x/\Delta$ and $J_\nu(z)$ being the $\nu$-order Bessel function of the first kind.
Where, the bare tunnelling strengths along $x$- and $y$-directions are denoted as $t_x$ and $t_y$ respectively.

By using the WS-W basis, the field operators can be expanded as,
\begin{equation}
  \hatd{\psi}_{\sigma}(\mbr)=\sum_{l,m}\phi_\sigma^*(\mbr-\mbr_{l,m})\hatd{a}_{l,m,\sigma},
\end{equation}
where $\hatd{a}_{l,m,\sigma}$ creates a $\sigma$-component boson at the $(l,m)$-th lattice site.
Thus the many-body Hamiltonian reads,
\begin{eqnarray}\label{Eq.Ham.2ndQ.S}
  &&\hat{H}=
  \sum_{l',m',l,m,\sigma}t_{l',m';l,m}^\sigma\hatd{a}_{l',m',\sigma}\hat{a}_{l,m,\sigma}+
  \sum_{l',m',l,m,\sigma}V_{l',m';l,m}^\sigma(t)\hatd{a}_{l',m',\sigma}\hat{a}_{l,m,\sigma} \nonumber \\
  &&+\sum_{l',m',l'_2,m'_2,l_2,m_2,l,m,\sigma_1,\sigma_2}
  U_{l',m',l'_2,m'_2;l_2,m_2,l,m}^{\sigma_1,\sigma_2}
  \hatd{a}_{l',m',\sigma_1}
  \hatd{a}_{l'_2,m'_2,\sigma_2}
  \hat{a}_{l_2,m_2,\sigma_2}
  \hat{a}_{l,m,\sigma_1}, \nonumber \\
\end{eqnarray}
with the parameters
\begin{equation}
  \left\{\begin{array}{l}
  t_{l',m';l,m}^\sigma=\mathlarger{\int}\dif^2\mbr\phi_\sigma^*(\mbr-\mbr_{l',m'})
  \left(\hat{h}_{x,\sigma}+\hat{h}_y\right)\phi_\sigma(\mbr-\mbr_{l,m}) \\
  V_{l',m';l,m}^\sigma(t)=\mathlarger{\int}\dif^2\mbr\phi_\sigma^*(\mbr-\mbr_{l',m'})
  V_K(\mbr,t)\phi_\sigma(\mbr-\mbr_{l,m}) \\
  U_{l',m',l'_2,m'_2;l_2,m_2,l,m}^{\sigma_1,\sigma_2}=
  \mathlarger{\frac{1}{2}}g_{\sigma_1\sigma_2}\mathlarger{\int}\dif^2\mbr\big[
  \phi_{\sigma_1}^*(\mbr-\mbr_{l',m'})
  \phi_{\sigma_2}^*(\mbr-\mbr_{l'_2,m'_2}) \\
  \quad\quad\quad\quad\quad\quad\quad\quad
  \times\phi_{\sigma_2}(\mbr-\mbr_{l_2,m_2})
  \phi_{\sigma_1}(\mbr-\mbr_{l,m})\big]. \\
  \end{array}\right.
\end{equation}

Under the single-band tight-binding (SBTB) approximation, we have
\begin{eqnarray}
  &&t_{l',m';l,m}^\sigma=\alpha_\sigma\Delta{l}\delta^{l'}_{l}\delta^{m'}_{m}
  +t^y\delta^{l'}_{l}(\delta^{m'}_{m+1}+\delta^{m'}_{m-1}), \label{Eq.Ham.2ndQ.t.S} \\
  &&U_{l',m',l'_2,m'_2;l_2,m_2,l,m}^{\sigma_1,\sigma_2}=
  \frac{1}{2}U_{\sigma_1\sigma_2}
  \delta^{l'}_{l}\delta^{m'}_{m}\delta^{l'_2}_{l}
  \delta^{m'_2}_{m}\delta^{l_2}_{l}\delta^{m_2}_{m}, \label{Eq.Ham.2ndQ.U.S}
\end{eqnarray}
with
\begin{equation}
  U_{\sigma_1\sigma_2}=g_{\sigma_1\sigma_2}
  \int\dif x\left|\phi^{\mathrm{ws}x}_{\sigma_1}(x)\phi^{\mathrm{ws}x}_{\sigma_2}(x)\right|^2
  \int\dif y\left|\phi^{\mathrm{w}y}(y)\right|^4.
\end{equation}
The matrix elements of $V_K(\mbr,t)$ are given as,
\begin{eqnarray}
  V_{l',m';l,m}^\sigma(t)&=&\Omega{\int}\dif x{\int}\dif y\Big[
  \phi^{\mathrm{ws}x*}_{\sigma}(x)\phi^{\mathrm{ws}x}_{\sigma}(x-x_{l-l'})
  \phi^{\mathrm{w}y*}(y)\phi^{\mathrm{w}y}(y-y_{m-m'}) \nonumber \\
  &&\times \cos(k'_xx+k'_yy-\theta_{l',m'})\Big],
\end{eqnarray}
with $\theta_{l',m'}=\omega t-\phi_{l',m'}$,
and $\phi_{l',m'}=l'\phi_x+m'\phi_y$ with $\phi_x=k'_xd_x$ and $\phi_y=k'_yd_y$.
Define
\begin{equation}
  \left\{\begin{array}{l}
  I^{x,\cos}_{\sigma,l-l'}=\mathlarger{\int}\dif x\phi^{\mathrm{ws}x*}_\sigma(x)\phi^{\mathrm{ws}x}_\sigma(x-x_{l-l'})\cos(k'_xx) \\
  I^{x,\sin}_{\sigma,l-l'}=\mathlarger{\int}\dif x\phi^{\mathrm{ws}x*}_\sigma(x)\phi^{\mathrm{ws}x}_\sigma(x-x_{l-l'})\sin(k'_xx) \\
  I^{y,\cos}_{m-m'}=\mathlarger{\int}\dif y\phi^{\mathrm{w}y*}(y)\phi^{\mathrm{w}y}(y-y_{m-m'})\cos(k'_yy) \\
  I^{y,\sin}_{m-m'}=\mathlarger{\int}\dif y\phi^{\mathrm{w}y*}(y)\phi^{\mathrm{w}y}(y-y_{m-m'})\sin(k'_yy) \\
  \end{array}\right.,
\end{equation}
as $\cos(k'_xx+k'_yy-\theta_{l',m'})$ $=$
$\cos(k'_xx)\cos(k'_yy)\cos(\theta_{l',m'})$ $+$
$\cos(k'_xx)\sin(k'_yy)\sin(\theta_{l',m'})$ $+$
$\sin(k'_xx)\cos(k'_yy)\sin(\theta_{l',m'})$ $-$
$\sin(k'_xx)\sin(k'_yy)\cos(\theta_{l',m'})$,
we have
\begin{equation}
  \left\{\begin{array}{l}
  I^{x,\cos}_{\sigma,l-l'}=\delta^{l'}_{l}I^{x,\cos}_{\sigma,0}+\delta^{l'}_{l-1}I^{x,\cos}_{\sigma,1}+\delta^{l'}_{l+1}I^{x,\cos}_{\sigma,-1} \\
  I^{x,\sin}_{\sigma,l-l'}=\delta^{l'}_{l}I^{x,\sin}_{\sigma,0}+\delta^{l'}_{l-1}I^{x,\sin}_{\sigma,1}+\delta^{l'}_{l+1}I^{x,\sin}_{\sigma,-1} \\
  I^{y,\cos}_{m-m'}=\delta^{m'}_{m}I^{y,\cos}_{0} \\
  I^{y,\sin}_{m-m'}=\delta^{m'}_{m}I^{y,\sin}_{0} \\
  \end{array}\right..
\end{equation}

There are several different types of Wannier functions, it is better to use the maximally localized Wannier functions for constructing $\phi^{\mathrm{w}x}(x-x_j)$ and $\phi^{\mathrm{w}y}(y-y_m)$.
The symmetry of the lattice potential implies the symmetric nature of the maximally localized Wannier functions~\cite{Kohn1959}
(i.e., they are either symmetric or antisymmetric).
Therefore, under the SBTB approximation, we have the following identities: $I^{y,\sin}_{0}=0$, $I^{x,\sin}_{\sigma,0}=0$, and
$I^{x,\cos}_{\sigma,1}\cos(\theta_{l,m})+I^{x,\sin}_{\sigma,1}\sin(\theta_{l,m})
=I^{x,\cos}_{\sigma,-1}\cos(\theta_{l+1,m})+I^{x,\sin}_{\sigma,-1}\sin(\theta_{l+1,m})$.
As $I^{x,\cos}_{\sigma,0}$ is $\sigma$-independent,
one can define $I^x_0=I^{x,\cos}_{\sigma,0}$ and $I^y_0=I^{y,\cos}_{0}$, therefore one can obtain
\begin{eqnarray}\label{Eq.Ham.2ndQ.V.S}
  V_{l',m';l,m}^\sigma=\Omega I^y_0\delta^{m'}_{m}\Big\{&&\delta^{l'}_{l}I^x_0\cos(\theta_{l,m}) \nonumber \\
  &&+\delta^{l'}_{l-1}\big[I^{x,\cos}_{\sigma,1}\cos(\theta_{l-1,m})+I^{x,\sin}_{\sigma,1}\sin(\theta_{l-1,m})\big] \nonumber \\
  &&+\delta^{l'}_{l+1}\big[I^{x,\cos}_{\sigma,-1}\cos(\theta_{l+1,m})+I^{x,\sin}_{\sigma,-1}\sin(\theta_{l+1,m})\big]\Big\}.
\end{eqnarray}

From equations (\ref{Eq.Ham.2ndQ.S}),~(\ref{Eq.Ham.2ndQ.t.S}),~(\ref{Eq.Ham.2ndQ.U.S}),~and~(\ref{Eq.Ham.2ndQ.V.S}),
the SBTB Hamiltonian can be written as,
\begin{equation}
  \hat{H}=\hat{H}^\mathrm{D}+\hat{H}^{\mathrm{OD}x}+\hat{H}^{\mathrm{OD}y}+\hat{H}^\mathrm{DI},
\end{equation}
with
\begin{eqnarray}
  &&\hat{H}^\mathrm{D}
  =\sum_{l,m,\sigma}\bigl[\alpha_\sigma\Delta l
  +\Omega I^x_0I^y_0\cos(\theta_{l,m})\bigr]\hat{n}_{l,m,\sigma}, \\
  &&\hat{H}^{\mathrm{OD}x}=\sum_{l,m,\sigma}\Omega
  I^y_0I^{x}_{l,m,\sigma}\left(\hatd{a}_{l+1,m,\sigma}\hat{a}_{l,m,\sigma}
  +\mathrm{h.c.}\right), \\
  &&\hat{H}^{\mathrm{OD}y}=
  \sum_{l,m,\sigma}t^y\left(\hatd{a}_{l,m+1,\sigma}\hat{a}_{l,m,\sigma}
  +\mathrm{h.c.}\right), \\
  &&\hat{H}^\mathrm{DI}=
  \sum_{l,m,\sigma_1,\sigma_2}\frac{1}{2}U_{\sigma_1\sigma_2}
  \hat{n}_{l,m,\sigma_1}\left(\hat{n}_{l,m,\sigma_2}
  -\delta_{\sigma_1,\sigma_2}\right),
\end{eqnarray}
where $\hat{n}_{l,m,\sigma}=\hatd{a}_{l,m,\sigma}\hat{a}_{l,m,\sigma}$
and $I^{x}_{l,m,\sigma}=I^{x,\cos}_{\sigma,1}\cos(\theta_{l,m})
+I^{x,\sin}_{\sigma,1}\sin(\theta_{l,m})$.

The time dependence of the diagonal term $\hat{H}^\mathrm{D}$ can be eliminated via a unitary transformation,
\begin{equation}
  \hat{U}=\exp\left(i\sum\nolimits_{l,m,\sigma}
  \Lambda_{l,m,\sigma}\hat{n}_{l,m,\sigma}\right),
\end{equation}
where,
\begin{equation}
  \Lambda_{l,m,\sigma}=-\alpha_\sigma l\omega t
  -\frac{\Omega}{\hbar\omega}I^x_0I^y_0\sin(\theta_{l,m})+l\theta_\sigma
\end{equation}
is real and time-dependent.
For convenience, we introduce a spin-dependent phase $\theta_\sigma$ whose value will be determined below.
The Hamiltonian in the rotating frame is given as $\hat{H}'=\hatd{U}\hat{H}\hat{U}-i\hbar\hatd{U}(\partial_t\hat{U})$.
For a resonant driving (i.e. $\hbar\omega=\Delta$),
we have $\hatd{U}\hat{H}^\mathrm{D}\hat{U}-i\hbar\hatd{U}(\partial_t\hat{U})=0$.
Thus, $\hat{H}'$ becomes as
\begin{equation}
  \hat{H}'=\hatd{U}\hat{H}^{\mathrm{OD}x}\hat{U}+\hatd{U}\hat{H}^{\mathrm{OD}y}\hat{U}+\hat{H}^\mathrm{DI}.
\end{equation}
Using the bosonic identity $e^{-i\theta\hat{n}}\hatd{a}e^{i\theta\hat{n}}=e^{-i\theta}\hatd{a}$,
we have
$\hatd{U}\hatd{a}_{l,m,\sigma}\hat{U}
=e^{-i\Lambda_{l,m,\sigma}}\hatd{a}_{l,m,\sigma}$
and
$\hatd{U}\hat{a}_{l,m,\sigma}\hat{U}
=e^{i\Lambda_{l,m,\sigma}}\hat{a}_{l,m,\sigma}$.
Consequently, one can find
\begin{eqnarray}
  &&\hatd{U}\hatd{a}_{l+1,m,\sigma}\hat{a}_{l,m,\sigma}\hat{U}
  =e^{i(\Lambda_{l,m,\sigma}
  -\Lambda_{l+1,m,\sigma})}\hatd{a}_{l+1,m,\sigma}\hat{a}_{l,m,\sigma},
  \\
  &&\hatd{U}\hatd{a}_{l,m+1,\sigma}\hat{a}_{l,m,\sigma}\hat{U}
  =e^{i(\Lambda_{l,m,\sigma}
  -\Lambda_{l,m+1,\sigma})}\hatd{a}_{l,m+1,\sigma}\hat{a}_{l,m,\sigma},
\end{eqnarray}
with the time-dependent phases
\begin{eqnarray}
  &&\Lambda_{l,m,\sigma}-\Lambda_{l+1,m,\sigma}
  =\alpha_\sigma\omega t-\theta_\sigma
  -\Gamma_x\cos
  (\omega t-\phi_{l,m}-\frac{\phi_x}{2}), \\
  &&\Lambda_{l,m,\sigma}-\Lambda_{l,m+1,\sigma}=-\Gamma_y\cos(\omega t-\phi_{l,m}-\frac{\phi_y}{2}).
\end{eqnarray}
Here,
$\Gamma_\alpha=\frac{2\Omega}{\hbar\omega}
I^x_0I^y_0\sin(\frac{1}{2}\phi_\alpha)$
with $\alpha=x$ and $y$.
Using the variant of the Jacobi-Anger identity,
$e^{-iz\cos(\theta)}=e^{iz\sin(\theta-\frac{\pi}{2})}=\sum_rJ_r(z)e^{ir(\theta-\frac{\pi}{2})}$,
the phase factors are given as
\begin{eqnarray}
  e^{i(\Lambda_{l,m,\sigma}-\Lambda_{l+1,m,\sigma})}
  &=&\sum\nolimits_r\big[J_r(\Gamma_x)
  e^{i(\alpha_\sigma+r)\omega t}
  \times e^{-ir(\phi_{l,m}+\frac{\phi_x}{2}
  +\frac{\pi}{2})-i\theta_\sigma}\big], \\
  e^{i(\Lambda_{l,m,\sigma}-\Lambda_{l,m+1,\sigma})}
  &=&\sum\nolimits_rJ_r(\Gamma_y)
  e^{ir\omega t-ir(\phi_{l,m}+\frac{\phi_y}{2}+\frac{\pi}{2})}.
\end{eqnarray}
Therefore the off-diagonal terms of the Hamiltonian $\hat{H}'$ become as
\begin{eqnarray}
  \hatd{U}\hat{H}^{\mathrm{OD}x}\hat{U}=
  \sum\nolimits_{l,m,\sigma}\bigl[K^\sigma_{l,m}(t)\hatd{a}_{l+1,m,\sigma}\hat{a}_{l,m,\sigma}
  +\mathrm{h.c.}\bigr] \\
  \hatd{U}\hat{H}^{\mathrm{OD}y}\hat{U}=
  \sum\nolimits_{l,m,\sigma}\bigl[J_{l,m}(t)\hatd{a}_{l,m+1,\sigma}\hat{a}_{l,m,\sigma}
  +\mathrm{h.c.}\bigr]
\end{eqnarray}
with
\begin{eqnarray}
  K^\sigma_{l,m}(t)&=&\Omega I^y_0I^{x}_{l,m,\sigma}\sum\nolimits_r\big[J_r(\Gamma_x)
  e^{i(\alpha_\sigma+r)\omega t}
  e^{-ir(\phi_{l,m}+\frac{\phi_x}{2}+\frac{\pi}{2})-i\theta_\sigma}\big] \\
  J_{l,m}(t)&=&t^y\sum\nolimits_rJ_r(\Gamma_y)e^{ir\omega t-ir(\phi_{l,m}+\frac{\phi_y}{2}+\frac{\pi}{2})}.
\end{eqnarray}
Time-averaging over a period of $2\pi/\omega$ and using the identity
$\frac{1}{2\pi/\omega}\int^{2\pi/\omega}_{0}\dif{t}\,e^{ir\omega t}=\delta_{r,0}$ (for any integer $r$), one can obtain
\begin{equation}
  \left\{\begin{array}{l}
  \frac{1}{2\pi/\omega}\int^{2\pi/\omega}_{0}\dif{t}\,K^\sigma_{l,m}(t)
  =e^{i\alpha_\sigma\phi_{l,m}}\tilde{K}^\sigma, \\
  \frac{1}{2\pi/\omega}\int^{2\pi/\omega}_{0}\dif{t}\,J_{l,m}(t)=J.
  \end{array}
  \right.
\end{equation}
Here, $J=t^yJ_0(\Gamma_y)$ and the $\sigma$-dependent constant $\tilde{K}^\sigma$ is given as
\begin{eqnarray}
  \tilde{K}^\sigma&=&\frac{1}{2}\Omega I^y_0e^{-i\theta_\sigma}\bigl[
  (I^{x,\exp}_{\sigma,1})^*J_{1+\alpha_\sigma}
  (\Gamma_x)e^{i\frac{1}{2}(\alpha_\sigma+1)(\phi_x+\pi)} \nonumber \\
  &&+I^{x,\exp}_{\sigma,1}J_{1-\alpha_\sigma}
  (\Gamma_x)e^{i\frac{1}{2}(\alpha_\sigma-1)(\phi_x+\pi)}
\bigr]
\end{eqnarray}
with the notation
$I^{x,\exp}_{\sigma,1}=I^{x,\cos}_{\sigma,1}+iI^{x,\sin}_{\sigma,1}$.
Notice that $(I^{x,\exp}_{\uparrow,1})^*=e^{-i\phi_x}I^{x,\exp}_{\downarrow,1}$,
if we define
\begin{equation}
  \frac{1}{2}\Omega I^y_0[I^{x,\exp}_{\uparrow,1}J_0(\Gamma_x)-I^{x,\exp}_{\downarrow,1}J_2(\Gamma_x)]\equiv Ke^{i\theta_K}
\end{equation}
with $K>0$ and $\theta_K\in(-\pi,\pi]$, we have
\begin{equation}
  \left\{\begin{array}{l}
  \tilde{K}^\uparrow=Ke^{i(-\theta_\uparrow+\theta_K)} \\
  \tilde{K}^\downarrow=Ke^{-i(\theta_\downarrow+\phi_x-\theta_K)}
  \end{array}\right..
\end{equation}
The undetermined phases $\theta_{\sigma}$ are thus given as $\theta_\uparrow=\theta_K$ and $\theta_\downarrow=\theta_K-\phi_x$ such that
$\tilde{K}^\uparrow=\tilde{K}^\downarrow=K>0$.
Thus the effective Hamiltonian in the rotating frame
$\hat{H}_\mathrm{eff}=\frac{1}{2\pi/\omega}\int^{2\pi/\omega}_{0}\dif{t}\,\hat{H}'$ is given as
\begin{eqnarray}
  \hat{H}_\mathrm{eff}=&&\sum_{l,m,\sigma}
  \left(Ke^{i\alpha_\sigma\phi_{l,m}}
  \hatd{a}_{l+1,m,\sigma}\hat{a}_{l,m,\sigma}
  +\mathrm{h.c.}\right)
  +\sum_{l,m,\sigma}
  \left(J\hatd{a}_{l,m+1,\sigma}\hat{a}_{l,m,\sigma}
  +\mathrm{h.c.}\right) \nonumber \\
  &&+\sum_{l,m,\sigma_1,\sigma_2}\frac{1}{2}
  U_{\sigma_1\sigma_2}
  \hat{n}_{l,m,\sigma_1}\left(\hat{n}_{l,m,\sigma_2}
  -\delta_{\sigma_1,\sigma_2}\right).
\end{eqnarray}

Through a time-independent unitary transformation,
\begin{equation}
  \hat{U}'=\exp\left(i\sum\nolimits_{l,m,\sigma}
  \Lambda'_{l,m,\sigma}\hat{n}_{l,m,\sigma}\right),
\end{equation}
where
$\Lambda'_{l,m,\sigma}=\alpha_\sigma[\frac{1}{2}\phi_xl^2
-(\pi+\frac{1}{2}\phi_x)l]-m\pi$,
one can change the sign of $K$ and $J$.
Since
$\hat{U}'^\dagger\hatd{a}_{l,m,\sigma}\hat{U}'=
e^{-i\Lambda'_{l,m,\sigma}}\hatd{a}_{l,m,\sigma}$
and
$\hat{U}'^\dagger\hat{a}_{l,m,\sigma}\hat{U}'=
e^{i\Lambda'_{l,m,\sigma}}\hat{a}_{l,m,\sigma}$,
we have
\begin{eqnarray}
  &&\hat{U}'^\dagger\hatd{a}_{l+1,m,\sigma}\hat{a}_{l,m,\sigma}\hat{U}'
  =e^{i(\Lambda'_{l,m,\sigma}-\Lambda'_{l+1,m,\sigma})}
  \hatd{a}_{l+1,m,\sigma}\hat{a}_{l,m,\sigma}, \\
  &&\hat{U}'^\dagger\hatd{a}_{l,m+1,\sigma}\hat{a}_{l,m,\sigma}\hat{U}'
  =e^{i(\Lambda'_{l,m,\sigma}-\Lambda'_{l,m+1,\sigma})}
  \hatd{a}_{l,m+1,\sigma}\hat{a}_{l,m,\sigma},
\end{eqnarray}
with the phases
\begin{eqnarray}
  &&\Lambda'_{l,m,\sigma}-\Lambda'_{l+1,m,\sigma}
  =\alpha_\sigma(\pi-l\phi_x), \\
  &&\Lambda'_{l,m,\sigma}-\Lambda'_{l,m+1,\sigma}
  =\pi.
\end{eqnarray}
Thus the effective Hamiltonian becomes
\begin{eqnarray}\label{Eq.Bose-Hubbard.S}
  \hat{H}_\mathrm{B}&=&-\sum_{l,m,\sigma}\left[Ke^{i\alpha_\sigma m\Phi}
  \hatd{a}_{l+1,m,\sigma}\hat{a}_{l,m,\sigma}
  +\mathrm{h.c.}\right]
  -\sum_{l,m,\sigma}\left[J\hatd{a}_{l,m+1,\sigma}\hat{a}_{l,m,\sigma}
  +\mathrm{h.c.}\right] \nonumber \\
  &&+\sum_{l,m,\sigma_1,\sigma_2}\frac{1}{2}U_{\sigma_1\sigma_2}
  \hat{n}_{l,m,\sigma_1}\left(\hat{n}_{l,m,\sigma_2}
  -\delta_{\sigma_1,\sigma_2}\right)
\end{eqnarray}
with $\Phi=\phi_y=k'_yd_y$.
The above Hamiltonian is an interacting spinor Hofstadter model.

\section{Derivation of the generalized Heisenberg XXZ model}

In the strongly interacting regime [any of $(U_{\uparrow\uparrow},U_{\uparrow\downarrow},U_{\downarrow\downarrow})$ is far larger than any of $(K,J)$],
the model~(\ref{Eq.Bose-Hubbard.S}) with unity filling can be mapped onto a spin model, which is equivalent to a hard-core Bose-Hubbard model.
Below, by using the perturbation theory for degenerated many-body quantum systems~\cite{Takahashi1977},
we analytically derive an effective spin model up to second-order perturbation.

In the strongly interacting regime,
one can treat the hopping terms
\begin{equation}
  \hat{H}_1=\hat{H}_K+\hat{H}_J
\end{equation}
as a perturbation to the interaction term
\begin{equation}
  \hat{H}_0=\sum_{l,m,\sigma_1,\sigma_2}\frac{1}{2}U_{\sigma_1\sigma_2}
  \hat{n}_{l,m,\sigma_1}\left(\hat{n}_{l,m,\sigma_2}-\delta_{\sigma_1,\sigma_2}\right),
\end{equation}
where, the hopping terms are given as
\begin{equation}
  \left\{\begin{array}{l}
  \hat{H}_K=-K\sum_{l,m}\left[\hat{T}^{x}_{l,m}+\hat{T}^{x\dagger}_{l,m}\right] \\
  \hat{H}_J=-J\sum_{l,m}\left[\hat{T}^{y}_{l,m}+\hat{T}^{y\dagger}_{l,m}\right] \\
  \end{array}\right.
\end{equation}
with
\begin{equation}
  \left\{\begin{array}{l}
  \hat{T}^{x}_{l,m}=\sum_{\sigma}e^{i\alpha_\sigma m\Phi}
  \hatd{a}_{l+1,m,\sigma}\hat{a}_{l,m,\sigma} \\
  \hat{T}^{y}_{l,m}=\sum_{\sigma}
  \hatd{a}_{l,m+1,\sigma}\hat{a}_{l,m,\sigma} \\
  \end{array}\right..
\end{equation}
Obviously, any Fock state is an eigenstate of $\hat{H}_0$.
The Fock state for the system of unity filling is given as
\begin{equation}\label{Eq.Unity.Fill.Fock.States.S}
  \ket{\mathbf{n}}=\ket{\dots,n_{l,m,\sigma},\dots}
  =\prod_{l,m,\sigma}\frac{1}{\sqrt{n_{l,m,\sigma}!}}
  (\hatd{a}_{l,m,\sigma})^{n_{l,m,\sigma}}\ket{\mathbf{0}}
\end{equation}
where $\sum_{l,m,\sigma}n_{l,m,\sigma}=L_xL_y=L$
[$L_\alpha$ is the number of lattice sites along $\alpha$-direction ($\alpha=x$ and $y$), while $L$ is the total number of sites of the whole two-dimensional lattice]
and $\ket{\mathbf{0}}$ denotes the vacuum state.
According to the eigen-equation $\hat{H}_0\ket{\mathbf{n}}=E_\mathbf{n}\ket{\mathbf{n}}$, we have the eigen-energy,
\begin{equation}
  E_\mathbf{n}=\sum_{l,m,\sigma_1,\sigma_2}\frac{1}{2}
  U_{\sigma_1\sigma_2}n_{l,m,\sigma_1}
  \left(n_{l,m,\sigma_2}-\delta_{\sigma_1,\sigma_2}\right).
\end{equation}
Obviously, due to only one atom in each lattice site, the ground-state has energy $E_0=0$ and $2^L$-fold degeneracy.
In the Fock basis, the ground states are expressed as
\begin{equation}
  \ket{\mathbf{s}}=\ket{\dots,s_{l,m},\dots}=
  \prod_{l,m}\hatd{a}_{l,m,\sigma_{lm}}\ket{\mathbf{0}},
\end{equation}
with $s_{l,m}\in\{\uparrow,\downarrow\}$
and $\hat{H}_0\ket{\mathbf{s}}=E_0\ket{\mathbf{s}}$.

The projector onto the ground-state space $\mathcal{U}_0$ is,
\begin{equation}
  \hat{P}_0=\sum_{\mathbf{s}}\ket{\mathbf{s}}\bra{\mathbf{s}}.
\end{equation}
Introducing $\mathcal{V}_0$ as the orthogonal complement of $\mathcal{U}_0$,
the relevant projector onto $\mathcal{V}_0$ is,
\begin{equation}
  \hat{S}=-\sum_{E_\mathbf{n}\ne0}\frac{1}{E_\mathbf{n}}\ket{\mathbf{n}}\bra{\mathbf{n}}.
\end{equation}
Thus the effective Hamiltonian up to $2$nd order is given as,
\begin{equation}
  \hat{H}^{(2)}_{\mathrm{eff}}=\hat{P}_0\hat{H}_1\hat{S}\hat{H}_1\hat{P}_0.
\end{equation}
It's easy to find that $\hat{P}_0\hat{H}_K\hat{S}\hat{H}_J\hat{P}_0=\hat{P}_0\hat{H}_J\hat{S}\hat{H}_K\hat{P}_0=0$,
which gives
\begin{equation}\label{Eq.Heff.S}
  \hat{H}^{(2)}_{\mathrm{eff}}=\hat{P}_0\hat{H}_K\hat{S}\hat{H}_K\hat{P}_0+\hat{P}_0\hat{H}_J\hat{S}\hat{H}_J\hat{P}_0.
\end{equation}
Furthermore, since
\begin{equation}
  \left\{\begin{array}{l}
  \hat{P}_0\hat{T}^{\alpha}_{l,m}\hat{S}\hat{T}^{\alpha}_{l',m'}\hat{P}_0
  =\hat{P}_0\hat{T}^{\alpha\dagger}_{l,m}\hat{S}
  \hat{T}^{\alpha\dagger}_{l',m'}\hat{P}_0=0 \\
  \hat{P}_0\hat{T}^{\alpha}_{l,m}\hat{S}\hat{T}^{\alpha\dagger}_{l',m'}\hat{P}_0
  =\delta^{l'}_{l}\delta^{m'}_{m}\hat{P}_0\hat{T}^{\alpha}_{l,m}\hat{S}
  \hat{T}^{\alpha\dagger}_{l,m}\hat{P}_0 \\
  \hat{P}_0\hat{T}^{\alpha\dagger}_{l,m}\hat{S}\hat{T}^{\alpha}_{l',m'}\hat{P}_0
  =\delta^{l'}_{l}\delta^{m'}_{m}\hat{P}_0\hat{T}^{\alpha\dagger}_{l,m}\hat{S}
  \hat{T}^{\alpha}_{l,m}\hat{P}_0 \\
  \end{array}\right.
\end{equation}
for $\alpha=x$ and $y$, we have
\begin{equation}\label{Eq.P0HkSHkP0.S}
  \hat{P}_0\hat{H}_K\hat{S}\hat{H}_K\hat{P}_0=K^2\sum\limits_{l,m}
  (\hat{P}_0\hat{T}^{x}_{l,m}\hat{S}\hat{T}^{x\dagger}_{l,m}\hat{P}_0
  +\hat{P}_0\hat{T}^{x\dagger}_{l,m}\hat{S}\hat{T}^{x}_{l,m}\hat{P}_0),
\end{equation}
and
\begin{equation}\label{Eq.P0HjSHjP0.S}
  \hat{P}_0\hat{H}_J\hat{S}\hat{H}_J\hat{P}_0=J^2\sum\limits_{l,m}
  (\hat{P}_0\hat{T}^{y}_{l,m}\hat{S}\hat{T}^{y\dagger}_{l,m}\hat{P}_0
  +\hat{P}_0\hat{T}^{y\dagger}_{l,m}\hat{S}\hat{T}^{y}_{l,m}\hat{P}_0).
\end{equation}
This means that the effective Hamiltonian has two parts, which respectively correspond to the influences from the hopping terms of $x$- and $y$-directions.

As the effective Hamiltonian only involves the nearest-neighbor couplings,
it is sufficient to give its parameters by considering a system of two lattice sites.
For the hopping along $x$-direction, we take site-$(l,m)$ as site-$1$ and site-$(l+1,m)$ as site-$2$.
The two-site ground-states are
\begin{equation}
  \left\{\begin{array}{ll}
  \ket{\uparrow,\uparrow}=\hatd{a}_{1\uparrow}\hatd{a}_{2\uparrow}\ket{\mathbf{0}}, &
  \ket{\uparrow,\downarrow}=\hatd{a}_{1\uparrow}\hatd{a}_{2\downarrow}\ket{\mathbf{0}}, \\
  \ket{\downarrow,\uparrow}=\hatd{a}_{1\downarrow}\hatd{a}_{2\uparrow}\ket{\mathbf{0}}, &
  \ket{\downarrow,\downarrow}=\hatd{a}_{1\downarrow}\hatd{a}_{2\downarrow}\ket{\mathbf{0}}, \\
  \end{array}\right.
\end{equation}
with the eigenenergy $E_0=0$.
While the two-site excited states are
\begin{equation}
  \left\{\begin{array}{ll}
  \ket{\uparrow\uparrow,0}=\frac{1}{\sqrt{2}}(\hatd{a}_{1\uparrow})^2\ket{\mathbf{0}},&
  \ket{0,\uparrow\uparrow}=\frac{1}{\sqrt{2}}(\hatd{a}_{2\uparrow})^2\ket{\mathbf{0}},\\
  \ket{\downarrow\downarrow,0}=\frac{1}{\sqrt{2}}(\hatd{a}_{1\downarrow})^2\ket{\mathbf{0}},&
  \ket{0,\downarrow\downarrow}=\frac{1}{\sqrt{2}}(\hatd{a}_{2\downarrow})^2\ket{\mathbf{0}},\\
  \ket{\uparrow\downarrow,0}=\hatd{a}_{1\uparrow}\hatd{a}_{1\downarrow}\ket{\mathbf{0}},&
  \ket{0,\uparrow\downarrow}=\hatd{a}_{2\uparrow}\hatd{a}_{2\downarrow}\ket{\mathbf{0}},\\
  \end{array}\right.
\end{equation}
with eigenenergies $E_{\uparrow\uparrow,0}=E_{0,\uparrow\uparrow}=U_{\uparrow\uparrow}$,
$E_{\downarrow\downarrow,0}=E_{0,\downarrow\downarrow}=U_{\downarrow\downarrow}$
and $E_{\uparrow\downarrow,0}=E_{0,\uparrow\downarrow}=U_{\uparrow\downarrow}$.
Hence, we have the projectors
\begin{eqnarray}
  \hat{P}_0&=&\bigl[
  \hatd{a}_{1\uparrow}\hatd{a}_{2\uparrow}\ket{\mathbf{0}}\bra{\mathbf{0}}\hat{a}_{2\uparrow}\hat{a}_{1\uparrow}+
  \hatd{a}_{1\uparrow}\hatd{a}_{2\downarrow}\ket{\mathbf{0}}\bra{\mathbf{0}}\hat{a}_{2\downarrow}\hat{a}_{1\uparrow} \nonumber \\
  &&+
  \hatd{a}_{1\downarrow}\hatd{a}_{2\uparrow}\ket{\mathbf{0}}\bra{\mathbf{0}}\hat{a}_{2\uparrow}\hat{a}_{1\downarrow}+
  \hatd{a}_{1\downarrow}\hatd{a}_{2\downarrow}\ket{\mathbf{0}}\bra{\mathbf{0}}\hat{a}_{2\downarrow}\hat{a}_{1\downarrow}
  \bigr],\label{Eq.P0.in.2.Site.S}
\end{eqnarray}
\begin{eqnarray}
  \hat{S}&=&-\Big\{
  \frac{1}{2U_{\uparrow\uparrow}}\bigl[
  (\hatd{a}_{1\uparrow})^2\ket{\mathbf{0}}\bra{\mathbf{0}}(\hat{a}_{1\uparrow})^2+
  (\hatd{a}_{2\uparrow})^2\ket{\mathbf{0}}\bra{\mathbf{0}}(\hat{a}_{2\uparrow})^2\bigr] \nonumber \\
  &&+
  \frac{1}{2U_{\downarrow\downarrow}}\bigl[
  (\hatd{a}_{1\downarrow})^2\ket{\mathbf{0}}\bra{\mathbf{0}}(\hat{a}_{1\downarrow})^2+
  (\hatd{a}_{2\downarrow})^2\ket{\mathbf{0}}\bra{\mathbf{0}}(\hat{a}_{2\downarrow})^2\bigr] \nonumber \\
  &&+
  \frac{1}{U_{\uparrow\downarrow}}\bigl[
  \hatd{a}_{1\uparrow}\hatd{a}_{1\downarrow}\ket{\mathbf{0}}\bra{\mathbf{0}}\hat{a}_{1\downarrow}\hat{a}_{1\uparrow}+
  \hatd{a}_{2\uparrow}\hatd{a}_{2\downarrow}\ket{\mathbf{0}}\bra{\mathbf{0}}\hat{a}_{2\downarrow}\hat{a}_{2\uparrow}
  \Big\},
  \label{Eq.S.in.2.Site.S}
\end{eqnarray}
and
\begin{equation}\label{Eq.Tx.in.2.Site.S}
  \hat{T}^{x}_{l,m}=e^{im\Phi}\hatd{a}_{2\uparrow}\hat{a}_{1\uparrow}+
  e^{-im\Phi}\hatd{a}_{2\downarrow}\hat{a}_{1\downarrow}.
\end{equation}
Inserting equations~(\ref{Eq.P0.in.2.Site.S}),~(\ref{Eq.S.in.2.Site.S}),~and~(\ref{Eq.Tx.in.2.Site.S}) into equation~(\ref{Eq.P0HkSHkP0.S}),
and using the bosonic commutation relations and
the identity
$\hat{a}_{j,\sigma}\hatd{a}_{j',\sigma'}\ket{\mathbf{0}}
=\delta^{j}_{j'}\delta^{\sigma}_{\sigma'}\ket{\mathbf{0}}$
($j,~j'\in\{1,2\}$),
after some tedious algebra, we obtain
\begin{eqnarray}\label{Eq.P0TxSTxDP0.S}
  &&\hat{P}_0\hat{T}^{x}_{l,m}\hat{S}\hat{T}^{x\dagger}_{l,m}\hat{P}_0
  =\hat{P}_0\hat{T}^{x\dagger}_{l,m}\hat{S}\hat{T}^{x}_{l,m}\hat{P}_0
  \nonumber \\ 
  &&=-\Big[\frac{2}{U_{\uparrow\uparrow}}
  \hatd{a}_{1\uparrow}\hatd{a}_{2\uparrow}\hat{a}_{2\uparrow}\hat{a}_{1\uparrow}
  +\frac{2}{U_{\downarrow\downarrow}}
  \hatd{a}_{1\downarrow}\hatd{a}_{2\downarrow}\hat{a}_{2\downarrow}
  \hat{a}_{1\downarrow} \nonumber \\ 
  &&+\frac{1}{U_{\uparrow\downarrow}}(
  \hatd{a}_{1\downarrow}\hatd{a}_{2\uparrow}
  \hat{a}_{2\uparrow}\hat{a}_{1\downarrow}+
  \hatd{a}_{1\uparrow}\hatd{a}_{2\downarrow}
  \hat{a}_{2\downarrow}\hat{a}_{1\uparrow}) \nonumber \\ 
  &&+\frac{1}{U_{\uparrow\downarrow}}(
  e^{i2m\Phi}\hatd{a}_{1\downarrow}\hatd{a}_{2\uparrow}
  \hat{a}_{2\downarrow}\hat{a}_{1\uparrow}+
  e^{-i2m\Phi}\hatd{a}_{1\uparrow}\hatd{a}_{2\downarrow}
  \hat{a}_{2\uparrow}\hat{a}_{1\downarrow})\Big]. 
\end{eqnarray}
By introducing the pseudospin operators: $\hat{S}^+_{l,m}=\hatd{a}_{l,m,\uparrow}\hat{a}_{l,m,\downarrow}$,
$\hat{S}^-_{l,m}=\hatd{a}_{l,m,\downarrow}\hat{a}_{l,m,\uparrow}$,
and
$\hat{S}^z_{l,m}=\frac{1}{2}(\hat{n}_{l,m,\uparrow}-\hat{n}_{l,m,\downarrow})$
(we set $\hbar=1$ here and after),
equation~(\ref{Eq.P0TxSTxDP0.S}) can be rewritten as
\begin{eqnarray}
  &&\hat{P}_0\hat{T}^{x}_{l,m}\hat{S}\hat{T}^{x\dagger}_{l,m}\hat{P}_0
  +\hat{P}_0\hat{T}^{x\dagger}_{l,m}\hat{S}\hat{T}^{x}_{l,m}\hat{P}_0
  \nonumber \\ 
  &&=-\Big[2\frac{1}{U_{\uparrow\downarrow}}
  \Big(e^{i2m\Phi}\hat{S}^+_2\hat{S}^-_1+e^{-i2m\Phi}\hat{S}^+_1\hat{S}^-_2\Big)
  +4\Big(\frac{1}{U_{\uparrow\uparrow}}+\frac{1}{U_{\downarrow\downarrow}}
  -\frac{1}{U_{\uparrow\downarrow}}\Big)\hat{S}^z_1\hat{S}^z_2
  \nonumber \\ 
  &&+2\Big(\frac{1}{U_{\uparrow\uparrow}}-\frac{1}{U_{\downarrow\downarrow}}\Big)
  (\hat{S}^z_1+\hat{S}^z_2) 
  +\Big(\frac{1}{U_{\uparrow\uparrow}}+\frac{1}{U_{\downarrow\downarrow}}
  +\frac{1}{U_{\uparrow\downarrow}}\Big)\Big]. 
\end{eqnarray}
Extended to the lattice, that is $1\rightarrow(l,m)$ and $2\rightarrow(l+1,m)$, we have
\begin{eqnarray}\label{Eq.Heff.x.S}
  &&\hat{P}_0\hat{T}^{x}_{l,m}\hat{S}\hat{T}^{x\dagger}_{l,m}\hat{P}_0
  +\hat{P}_0\hat{T}^{x\dagger}_{l,m}\hat{S}\hat{T}^{x}_{l,m}\hat{P}_0
  \nonumber \\ 
  &&=-\Big[2\frac{1}{U_{\uparrow\downarrow}}
  \Big(e^{i2m\Phi}\hat{S}^+_{l+1,m}\hat{S}^-_{l,m}
  +e^{-i2m\Phi}\hat{S}^+_{l,m}\hat{S}^-_{l+1,m}\Big) \nonumber \\ 
  &&+4\Big(\frac{1}{U_{\uparrow\uparrow}}+\frac{1}{U_{\downarrow\downarrow}}
  -\frac{1}{U_{\uparrow\downarrow}}\Big)
  \hat{S}^z_{l,m}\hat{S}^z_{l+1,m}
  +2\Big(\frac{1}{U_{\uparrow\uparrow}}-\frac{1}{U_{\downarrow\downarrow}}\Big)
  \Big(\hat{S}^z_{l,m}+\hat{S}^z_{l+1,m}\Big) \nonumber \\ 
  &&+\Big(\frac{1}{U_{\uparrow\uparrow}}+\frac{1}{U_{\downarrow\downarrow}}
  +\frac{1}{U_{\uparrow\downarrow}}\Big)
  \Big]. 
\end{eqnarray}

For the hopping along $y$-direction, we take site-$(l,m)$ as site-$1$ and site-$(l,m+1)$ as site-$2$.
Similarly, up to the second-order perturbation, we obtain
\begin{eqnarray}\label{Eq.Heff.y.S}
  &&\hat{P}_0\hat{T}^{y}_{l,m}\hat{S}\hat{T}^{y\dagger}_{l,m}\hat{P}_0
  +\hat{P}_0\hat{T}^{y\dagger}_{l,m}\hat{S}\hat{T}^{y}_{l,m}\hat{P}_0
  \nonumber \\ 
  &&=-\Bigg[2\frac{1}{U_{\uparrow\downarrow}}
  \left(\hat{S}^+_{l,m+1}\hat{S}^-_{l,m}+\hat{S}^+_{l,m}\hat{S}^-_{l,m+1}\right)
  +4\left(\frac{1}{U_{\uparrow\uparrow}}+\frac{1}{U_{\downarrow\downarrow}}
  -\frac{1}{U_{\uparrow\downarrow}}\right)
  \hat{S}^z_{l,m}\hat{S}^z_{l,m+1} \nonumber \\ 
  &&+2\left(\frac{1}{U_{\uparrow\uparrow}}-\frac{1}{U_{\downarrow\downarrow}}\right)
  (\hat{S}^z_{l,m}+\hat{S}^z_{l,m+1})
  +\left(\frac{1}{U_{\uparrow\uparrow}}+\frac{1}{U_{\downarrow\downarrow}}
  +\frac{1}{U_{\uparrow\downarrow}}\right)\Bigg]. 
\end{eqnarray}

Introducing $J_x=\slfrac{2K^2}{U_{\uparrow\downarrow}}$, $J_y=\slfrac{2J^2}{U_{\uparrow\downarrow}}$,
$V_x=4K^2(\slfrac{1}{U_{\uparrow\uparrow}}+\slfrac{1}{U_{\downarrow\downarrow}}-\slfrac{1}{U_{\uparrow\downarrow}})$,
$V_y=4J^2(\slfrac{1}{U_{\uparrow\uparrow}}+\slfrac{1}{U_{\downarrow\downarrow}}-\slfrac{1}{U_{\uparrow\downarrow}})$,
and $B_0=4(K^2+J^2)(\slfrac{1}{U_{\uparrow\uparrow}}-\slfrac{1}{U_{\downarrow\downarrow}})$,
from equations (\ref{Eq.Heff.S}),~(\ref{Eq.P0HkSHkP0.S}),~(\ref{Eq.P0HjSHjP0.S}),~(\ref{Eq.Heff.x.S}),~and~(\ref{Eq.Heff.y.S}),
we get the effective Hamiltonian,
\begin{eqnarray}\label{Eq.Heisenberg.model.S}
  \hat{H}^{(2)}_\mathrm{eff}=&&-\sum_{l,m}
  \Big[J_xe^{i2m\Phi}\hat{S}^+_{l+1,m}\hat{S}^-_{l,m}+\mathrm{h.c.}\Big]
  -\sum_{l,m}\Big[J_y\hat{S}^+_{l,m+1}\hat{S}^-_{l,m}+\mathrm{h.c.}\Big]
  \nonumber \\
  &&-\sum_{l,m}\Big(V_x\hat{S}^z_{l,m}\hat{S}^z_{l+1,m}
  +V_y\hat{S}^z_{l,m}\hat{S}^z_{l,m+1}\Big)-B_0\sum_{l,m}\hat{S}^z_{l,m}.
\end{eqnarray}
Here, we have removed a constant energy shift: $-(K^2+J^2)(\slfrac{1}{U_{\uparrow\uparrow}} +\slfrac{1}{U_{\downarrow\downarrow}}+\slfrac{1}{U_{\uparrow\downarrow}})L_xL_y$.

According to the Matsubara-Matsuda mapping~\cite{Matsubara1956}: $\ket{\downarrow}\leftrightarrow\ket{0}$,
$\ket{\uparrow}\leftrightarrow\ket{1}$,
$\hat{S}^+_{l,m}\leftrightarrow\hatd{b}_{l,m}$,
$\hat{S}^-_{l,m}\leftrightarrow\hat{b}_{l,m}$,
and $\hat{S}^z_{l,m}\leftrightarrow(\hat{n}_{l,m}
-\frac{1}{2})\equiv(\hatd{b}_{l,m}\hat{b}_{l,m}-\frac{1}{2})$,
the magnon excitations can be described by hard-core bosons and so that the two-dimensional Heisenberg spin model~(\ref{Eq.Heisenberg.model.S}) is equivalent to a two-dimensional hard-core Bose-Hubbard model subjected to a synthetic gauge field,
\begin{eqnarray}
  \hat{H}_\mathrm{HC}&=&-J_x\sum_{l,m}
  \Big[\Big(e^{i2m\Phi}\hatd{b}_{l+1,m}\hat{b}_{l,m}
  +\lambda\hatd{b}_{l,m+1}\hat{b}_{l,m}\Big)+\mathrm{h.c.}\Big]
  \nonumber \\
  &&-V_x\sum_{l,m}\Big(\hat{n}_{l,m}\hat{n}_{l+1,m}
  +\lambda\hat{n}_{l,m}\hat{n}_{l,m+1}\Big)+\varepsilon_0\sum_{l,m}\hat{n}_{l,m}
\end{eqnarray}
with
$V_x=\Delta_x$, $V_y=\Delta_y$ and $\varepsilon_0=\Delta_x+\Delta_y-B_0$.
Here we have removed a constant energy shift
$\frac{1}{4}(2B_0-\Delta_x-\Delta_y)L_xL_y$.
Since the term $\varepsilon_0\sum_{l,m}\hat{n}_{l,m}$ commutes with the other part of the Hamiltonian,
it only causes a constant energy shift and thus can be removed from the Hamiltonian without changing the physics.
Finally, our effective hard-core boson model obeys,
\begin{eqnarray}\label{Eq.Hamiltonian.S}
  \hat{H}&=&-J_x\sum_{l,m}\Big[\Big(e^{i2\pi\beta m}
  \hatd{b}_{l+1,m}\hat{b}_{l,m}
  +\lambda\hatd{b}_{l,m+1}\hat{b}_{l,m}\Big)+\mathrm{h.c.}\Big] \nonumber \\
  &&-V_x\sum_{l,m}\Big(\hat{n}_{l,m}\hat{n}_{l+1,m}
  +\lambda\hat{n}_{l,m}\hat{n}_{l,m+1}\Big)
\end{eqnarray}
with $\beta=\Phi/\pi$ and $\lambda=J^2/K^2$.

\section{Derivation of the effective single-particle model for two-magnon bound-states}

By regarding a two-magnon bound state as a quasi-particle, we analytically derive an effective single-particle model via the Schrieffer-Wolff transformation~\cite{LossSWRev2011AoP}.
As bound-states appear when $|V_x/J_x|\gg1$, one can treat the hopping term
\begin{eqnarray}\label{Eq.Hop}
  \hat{H}_1&=&-J_x\sum_{l,m}\left[\left(e^{i2\pi\beta m}\hatd{b}_{l+1,m}\hat{b}_{l,m}
  +\lambda\hatd{b}_{l,m+1}\hat{b}_{l,m}\right)+\mathrm{h.c.}\right]
\end{eqnarray}
as a perturbation to the interaction term
\begin{equation}\label{Eq.Interaction}
  \hat{H}_0=-V_x\sum_{l,m}\Big(\hat{n}_{l,m}\hat{n}_{l+1,m}
  +\lambda\hat{n}_{l,m}\hat{n}_{l,m+1}\Big).
\end{equation}
Obviously, all two-magnon Fock states $\ket{l_1,m_1;l_2,m_2}=\hatd{b}_{l_1,m_1}\hatd{b}_{l_2,m_2}\ket{\mathbf{0}}$ are eigenstates of $\hat{H}_0$ with eigenvalues
$E_{l_1,m_1;l_2,m_2}=-V_x(\delta^{l_1\pm1,m_1}_{l_2,m_2}+\lambda\delta^{l_1,m_1\pm1}_{l_2,m_2})$.
The two-magnon bound-states can be approximated by superpositions: $\ket{G^x_{l\comma m}}=\ket{l,m;l+1,m}$ and $\ket{G^y_{l\comma m}}=\ket{l,m;l,m+1}$, which are also eigenstates of $\hat{H}_0$ with eigenvalues $E^x_0=-V_x$ and $E^y_0=-\lambda V_x$ (where $\hat{H}_0\ket{G^x_{l\comma m}}=E^x_0\ket{G^x_{l\comma m}}$ and $\hat{H}_0\ket{G^y_{l\comma m}}=E^y_0\ket{G^y_{l\comma m}}$).

Using SW transformation~\cite{LossSWRev2011AoP}, the effective single-particle Hamiltonian up to second-order reads
\begin{eqnarray}
  \hat{H}^{(2)}_\mathrm{eff}&=&\hat{h}_0+\hat{h}_2, \\
  \hat{h}_0&=&-V_x\left(\hat{P}_1+\lambda\hat{P}_2\right), \\
  \hat{h}_2&=&
  -\frac{1}{V_x}\left(\hat{P}_1\hat{H}_1\hat{H}_1\hat{P}_1
  +\frac{1}{\lambda}\hat{P}_2\hat{H}_1\hat{H}_1\hat{P}_2\right) \nonumber \\
  &&-\frac{\lambda+1}{2\lambda V_x}\left(
  \hat{P}_1\hat{H}_1\hat{H}_1\hat{P}_2
  +\hat{P}_2\hat{H}_1\hat{H}_1\hat{P}_1\right). \label{Eq.Ham.2nd.Eff}
\end{eqnarray}
Here, the two bound-state projectors are defined as
\begin{equation}
\left\{\begin{array}{l}
  \hat{P}_1=\sum\limits_{l,m}\ket{G^x_{l\comma m}}\bra{G^x_{l\comma m}},\\
  \hat{P}_2=\sum\limits_{l,m}\ket{G^y_{l\comma m}}\bra{G^y_{l\comma m}}.\\
\end{array}\right.
\end{equation}
For convenience, we introduce the following notations
\begin{eqnarray}
\left\{\begin{array}{l}
  \hat{H}_{J_x}=-J_x\sum_{l,m}
  \left(\hat{t}^x_{l,m}+\hat{t}^{x\dagger}_{l,m}\right),\\
  \hat{H}_{J_y}=-\lambda J_x\sum_{l,m}\left(\hat{t}^y_{l,m}+\hat{t}^{y\dagger}_{l,m}
  \right),\\
\end{array}\right.
\end{eqnarray}
with
\begin{eqnarray}
\left\{\begin{array}{l}
  \hat{t}^x_{l,m}=e^{i2\pi\beta m}\hatd{b}_{l+1,m}\hat{b}_{l,m},\\
  \hat{t}^y_{l,m}=\hatd{b}_{l,m+1}\hat{b}_{l,m}.\\
\end{array}\right.
\end{eqnarray}
It is easy to find that
\begin{equation}
  \left\{\begin{array}{l}
  \hat{P}_1\hat{H}_{J_x}\hat{H}_{J_y}\hat{P}_1=\hat{P}_1\hat{H}_{J_y}\hat{H}_{J_x}\hat{P}_1=0 \\
  \hat{P}_2\hat{H}_{J_x}\hat{H}_{J_y}\hat{P}_2=\hat{P}_2\hat{H}_{J_y}\hat{H}_{J_x}\hat{P}_2=0 \\
  \hat{P}_1\hat{H}_{J_x}\hat{H}_{J_x}\hat{P}_2=\hat{P}_1\hat{H}_{J_y}\hat{H}_{J_y}\hat{P}_2=0 \\
  \hat{P}_1\hat{H}_{J_x}\hat{H}_{J_y}\hat{P}_2=\hat{P}_2\hat{H}_{J_y}\hat{H}_{J_x}\hat{P}_1=0 \\
  \end{array}\right.
\end{equation}
As $\hat{H}_1=\hat{H}_{J_x}+\hat{H}_{J_y}$, we have
\begin{equation}\label{Eq.Ham.2nd.Eff.0}
  \left\{\begin{array}{l}
  \hat{P}_1\hat{H}_{1}\hat{H}_{1}\hat{P}_1=\hat{P}_1\hat{H}_{J_x}\hat{H}_{J_x}\hat{P}_1+\hat{P}_1\hat{H}_{J_y}\hat{H}_{J_y}\hat{P}_1 \\
  \hat{P}_2\hat{H}_{1}\hat{H}_{1}\hat{P}_2=\hat{P}_2\hat{H}_{J_x}\hat{H}_{J_x}\hat{P}_2+\hat{P}_2\hat{H}_{J_y}\hat{H}_{J_y}\hat{P}_2 \\
  \hat{P}_1\hat{H}_{1}\hat{H}_{1}\hat{P}_2=\hat{P}_1\hat{H}_{J_y}\hat{H}_{J_x}\hat{P}_2 \\
  \hat{P}_2\hat{H}_{1}\hat{H}_{1}\hat{P}_1=\hat{P}_2\hat{H}_{J_x}\hat{H}_{J_y}\hat{P}_1. \\
  \end{array}\right.
\end{equation}
By using the hard-core bosonic commutation relations, one can obtain
\begin{equation}
\left\{\begin{array}{lll}
  \hat{H}_{J_x}\ket{G^x_{l\comma m}}&=&-J_x\Big(
  e^{i2\pi\beta m}\hatd{b}_{l,m}\hatd{b}_{l+2,m}\\
  &&\quad\quad\quad
  +e^{-i2\pi\beta m}\hatd{b}_{l-1,m}\hatd{b}_{l+1,m}\Big)\ket{\mathbf{0}}\\
  \hat{H}_{J_x}\ket{G^y_{l\comma m}}&=&-J_x\Big(
  e^{i2\pi\beta m}\hatd{b}_{l,m+1}\hatd{b}_{l+1,m}\\
  &&\quad\quad\quad
  +e^{i2\pi\beta(m+1)}\hatd{b}_{l,m}\hatd{b}_{l+1,m+1}\\
  &&\quad\quad\quad
  +e^{-i2\pi\beta m}\hatd{b}_{l-1,m}\hatd{b}_{l,m+1}\\
  &&\quad\quad\quad
  +e^{-i2\pi\beta(m+1)}\hatd{b}_{l-1,m+1}\hatd{b}_{l,m}\Big)\ket{\mathbf{0}}\\
  \hat{H}_{J_y}\ket{G^x_{l\comma m}}&=&-\lambda J_x\Big(
  \hatd{b}_{l,m+1}\hatd{b}_{l+1,m}+\hatd{b}_{l,m}\hatd{b}_{l+1,m+1}\\
  &&\quad\quad\quad
  +\hatd{b}_{l,m-1}\hatd{b}_{l+1,m}
  +\hatd{b}_{l,m}\hatd{b}_{l+1,m-1}\Big)\ket{\mathbf{0}}\\
  \hat{H}_{J_y}\ket{G^y_{l\comma m}}&=&-\lambda J_x\Big(
  \hatd{b}_{l,m}\hatd{b}_{l,m+2}+\hatd{b}_{l,m-1}\hatd{b}_{l,m+1}\Big)\ket{\mathbf{0}}.\\
\end{array}\right.
\end{equation}
Therefore, we get
\begin{equation}\label{Eq.Ham.2nd.Eff.1}
\left\{\begin{array}{lcl}
  \hat{P}_1\hat{H}_{J_x}\hat{H}_{J_x}\hat{P}_1&=&J^2_x\sum\limits_{l,m}\Big(
  e^{i4\pi\beta m}\ket{G^x_{l+1\comma m}}\bra{G^x_{l\comma m}}+\mathrm{h.c.}\\
  &&\quad\quad\quad
  +2\ket{G^x_{l\comma m}}\bra{G^x_{l\comma m}}\Big)\\
  \hat{P}_1\hat{H}_{J_y}\hat{H}_{J_y}\hat{P}_1&=&2\lambda^2J^2_x\sum\limits_{l\comma m}\Big(
  \ket{G^x_{l\comma m+1}}\bra{G^x_{l\comma m}}+\mathrm{h.c.}\\
  &&\quad\quad\quad
  +2\ket{G^x_{l\comma m}}\bra{G^x_{l\comma m}}\Big)\\
  \hat{P}_2\hat{H}_{J_x}\hat{H}_{J_x}\hat{P}_2&=&2J^2_x\sum\limits_{l,m}\Big(
  e^{i2\pi\beta(2m+1)}\ket{G^y_{l+1\comma m}}\bra{G^y_{l\comma m}}\\
  &&\quad\quad\quad
  +\mathrm{h.c.}+2\ket{G^y_{l\comma m}}\bra{G^y_{l\comma m}}\Big)\\
  \hat{P}_2\hat{H}_{J_y}\hat{H}_{J_y}\hat{P}_2&=&\lambda^2J^2_x\sum\limits_{l,m}\Big(
  \ket{G^y_{l\comma m+1}}\bra{G^y_{l\comma m}}+\mathrm{h.c.}\\
  &&\quad\quad\quad
  +2\ket{G^y_{l\comma m}}\bra{G^y_{l\comma m}}\Big)\\
\end{array}\right.
\end{equation}
and
\begin{eqnarray}\label{Eq.Ham.2nd.Eff.2}
  &&\hat{P}_1\hat{H}_{J_y}\hat{H}_{J_x}\hat{P}_2+
  \hat{P}_2\hat{H}_{J_x}\hat{H}_{J_y}\hat{P}_1\nonumber\\
  &&=2\lambda J_x^2\cos(\pi\beta)\sum_{l,m}\Big[
  e^{i2\pi\beta m}e^{i\pi\beta}\Big(
  \ket{G^x_{l\comma m}}\bra{G^y_{l\comma m}}+\ket{G^y_{l+1\comma m}}\bra{G^x_{l\comma m}}\nonumber\\
  &&\quad\quad\quad\quad\quad\quad\quad
  +\ket{G^x_{l\comma m+1}}\bra{G^y_{l\comma m}}
  +\ket{G^y_{l+1\comma m}}\bra{G^x_{l\comma m+1}}
  \Big)+\mathrm{h.c.}\Big].
\end{eqnarray}
Insert equations~(\ref{Eq.Ham.2nd.Eff.0}),~(\ref{Eq.Ham.2nd.Eff.1}),~(\ref{Eq.Ham.2nd.Eff.2}) into Eq.~(\ref{Eq.Ham.2nd.Eff}), we obtain
\begin{eqnarray}\label{Eq.EffHam.G.S}
  \hat{H}^{(2)}_\mathrm{eff}=&-&J_\mathrm{eff}\sum_{l,m}\Big\{\Big[e^{i4\pi\beta m}
  \ket{G^x_{l+1\comma m}}\bra{G^x_{l\comma m}}
  +2\lambda^2
  \ket{G^x_{l\comma m+1}}\bra{G^x_{l\comma m}}\nonumber\\
  &&+\frac{2}{\lambda}e^{i4\pi\beta m}
  e^{i2\pi\beta}\ket{G^y_{l+1\comma m}}\bra{G^y_{l\comma m}}
  +\lambda
  \ket{G^y_{l\comma m+1}}\bra{G^y_{l\comma m}}\nonumber\\
  &&+J_{xy}e^{i2\pi\beta m}e^{i\pi\beta}\Big(
  \ket{G^x_{l\comma m}}\bra{G^y_{l\comma m}}
  +\ket{G^y_{l+1\comma m}}\bra{G^x_{l\comma m}}\nonumber\\
  &&+\ket{G^x_{l\comma m+1}}\bra{G^y_{l\comma m}}
  +\ket{G^y_{l+1\comma m}}\bra{G^x_{l\comma m+1}}
  \Big)+\mathrm{h.c.}\Big] \nonumber \\
  &&+\epsilon_x\ket{G^x_{l\comma m}}\bra{G^x_{l\comma m}}
  +\epsilon_y
  \ket{G^y_{l\comma m}}\bra{G^y_{l\comma m}}\Big\}.
\end{eqnarray}
Here, $J_\mathrm{eff}=J_x^2/V_x$, $J_{xy}=(\lambda+1)\cos(\pi\beta)$,
$\epsilon_x=V_x^2/J_x^2+2+4\lambda^2$,
and $\epsilon_y=\lambda V_x^2/J_x^2+2\lambda+4/\lambda$.

In order to capture the single-particle nature of the bound-states, we introduce the creation operators $\hatd{A}_{l,m}$ and $\hatd{B}_{l,m}$ as follows: $\hatd{A}_{l,m}$ creates a quasi-particle in the $x$-type bound-state $\ket{G^x_{l\comma m}}$, while $\hatd{B}_{l,m}$ creates a quasi-particle in the $y$-type bound-state $\ket{G^y_{l\comma m}}$.
That is, we define a mapping between two-magnon bound-states and single-particle states: $\ket{G^x_{l\comma m}}\Leftrightarrow\hatd{A}_{l,m}\ket{\mathbf{0}}$
and $\ket{G^y_{l\comma m}}\Leftrightarrow\hatd{B}_{l,m}\ket{\mathbf{0}}$.
Thus the effective single-particle Hamiltonian~(\ref{Eq.EffHam.G.S}) becomes
\begin{eqnarray}\label{Eq.Effective.Ham.S}
  \hat{H}_\mathrm{eff}=&-&J_\mathrm{eff}\sum_{lm}\Big\{\Big[e^{i4\pi\beta m}
  \hatd{A}_{l+1,m}\hat{A}_{l,m}+2\lambda^2\hatd{A}_{l,m+1}\hat{A}_{l,m} \nonumber \\
  &&+\frac{2}{\lambda}e^{i4\pi\beta m}e^{i2\pi\beta}\hatd{B}_{l+1,m}\hat{B}_{l,m}
  +\lambda\hatd{B}_{l,m+1}\hat{B}_{l,m} \nonumber \\
  &&+J_{xy}e^{i2\pi\beta m}e^{i\pi\beta}\Big(
  \hatd{A}_{l,m}\hat{B}_{l,m}+\hatd{B}_{l+1,m}\hat{A}_{l,m} \nonumber \\
  &&+\hatd{A}_{l,m+1}\hat{B}_{l,m}+\hatd{B}_{l+1,m}\hat{A}_{l,m+1}
  \Big)+\mathrm{h.c.}\Big] \nonumber \\
  &&+\epsilon_x\hatd{A}_{l,m}\hat{A}_{l,m}+\epsilon_y\hatd{B}_{l,m}\hat{B}_{l,m}\Big\},
\end{eqnarray}
which describes a Hofstadter superlattice with two coupled standard Hofstadter sublattices $A$ and $B$.

\end{document}